\documentclass{raa}
\usepackage{graphicx,times}
\usepackage{tikz,pgfbaselayers}
\usepackage{amsmath,amssymb,amsfonts}
\usepackage{pb-diagram}
\usepackage{hyperref}
\usepackage{natbib}

\makeatletter
\@namedef{dgt@nnne}{\dg@DX=1 \dg@DY=3 \dg@SIZE=1}
\makeatother

\begin{document}
\title{Accelerated direct demodulation method for image reconstruction with spherical data from Hard X-ray Modulation Telescope}

\volnopage{Vol.0 (200x) No.0, 000--000}      
\setcounter{page}{1}                         

\author{Z.-X. Huo\inst{1,2,3} \and J.-F. Zhou\inst{1,2,3}}
\institute{Department of Engineering Physics and Center for Astrophysics, Tsinghua University, Beijing 100084, China; {\it zhoujf@tsinghua.edu.cn}\\
\and
Key Laboratory of Particle \& Radiation Imaging (Tsinghua University), Ministry of Education, Beijing 100084, China\\
\and
Key Laboratory of High Energy Radiation Imaging Fundamental Science for National Defense, Beijing 100084, China}
\date{}
\abstract{
The hard X-ray modulation telescope mission HXMT is mainly devoted to performing an all-sky survey at $1\;\mathrm{keV}$ -- $250\;\mathrm{keV}$ with both high sensitivity and high spatial resolution. The observed data reduction as well as the image reconstruction for HXMT can be achieved by direct demodulation method (DDM). However the original DDM is computationally too expensive for multi-dimensional data with high resolution to employ for HXMT data. In this article we propose an accelerated direct demodulation method adapted for data from HXMT. Simulations are also presented to demonstrate this method.
\keywords{methods: data analysis --- methods: numerical --- techniques: image processing --- instrumentation: high angular resolution}
}
\authorrunning{Z.-X. Huo \& J.-F. Zhou}
\titlerunning{Accelerated DDM for HXMT}
\maketitle

\section{Introduction}
\subsection{Hard X-ray modulation telescope}
The hard X-ray modulation telescope (HXMT) is a low orbit space telescope runs at a circular orbit with a $43^\circ$ inclinatioin at the altitude of $550\;\mathrm{km}$ \citep{li2007,lu2010}.
There are three detectors on board, including \emph{High energy X-ray detector} (HE), \emph{medium energy X-ray detector} (ME), and \emph{low energy X-ray detector} (LE).
HE is the primary one, which consists of 18 modules and each module is composed of an HE collimator, a phoswich scintillation detector and readout electronics \citep{li2007,han2010}.

The HE collimator is used to define the field of view (FOV) of the detector\citep{han2010}.
There are two types of HE collimators differ in their FOVs: 15 of them are with $5.7^\circ \times 1^\circ$ FOVs, and 3 of them are with $5.7^\circ \times 5.7^\circ$ FOVs.
Their optical axes are parallel but the directions of their cross-sections are different.

There are two imaging observation modes designed for the scientific goal: the all-sky survey mode as well as the deep imaging observations of selected sky regions.
The all-sky survey is performed through the orbiting of the satellite as well as the precession of its orbit plane\citep{li2007}.
The orbit period is about $95.5\;\mathrm{min}$ while the precession rate is $-5.45^\circ$ per day\citep{lu2010}.
The deep imaging observations will be performed by pointed observations with pointing directions distributed uniformly in the region and the pointed observations will be performed by progressive scanning.

\subsection{Modulation}
According to \citet{li1994}, an observation process can be modeled as
\begin{equation}
\int p(\omega,x) f(x) \mathrm{d}x = d(\omega)\text{,}
\label{eq-modulate}
\end{equation}
where $f(x)$ is the intensity distribution of the object (i.e., the image), $d(\omega)$ is the observed data modulated by integral kernel (modulation function) $p(\omega,x)$.
In practice the observed data is recorded as discrete trunks (3-D), maps (2-D) or series (1-D).
We have
\begin{equation}
d_k = \sum_{i=1}^N p_{k,i}f_{i}\text{,}\;\forall k=1,\cdots,M\text{,}
\label{eq-modulate-disc}
\end{equation}
or
\begin{equation}
\boldsymbol{d} = \boldsymbol{P}\boldsymbol{f}\text{,}
\label{eq-modulate-matrix}
\end{equation}
where $\boldsymbol{d}$ is an $M \times 1$ column vector (data vector) in data space $\{\boldsymbol{d}\}$ representing all possible observed data records, $\boldsymbol{f}$ is an $N \times 1$ column vector (object vector) in image space $\{\boldsymbol{f}\}$ representing all possible images, thus $\boldsymbol{P}$ is an $M \times N$ matrix, i.e., the \textbf{kernel matrix} in $\{\boldsymbol{d}\}\times\{\boldsymbol{f}\}$.

\subsection{Challenges in image reconstruction for HXMT}
Generally speaking, the direct demodulation method (DDM) is designed to solve the demodulation problem, i.e., finding a best object $f(x)$ to satisfy Eq. \ref{eq-modulate} while the kernel $p(\omega,x)$ and the observed data $d(\omega)$ are known\citep{li1993}.
Richardson--Lucy iteration\citep{richardson1972,lucy1974} is used for DDM, hence the numerical evaluation of modulations is the most time-consuming part.
The computational complexity increases rapidly as the size of the data increases in multi-dimensional modulation evaluation, as a result it is not feasible to reconstruct the all-sky image from HXMT observed data by the original DDM.
Hence accelerations of the method are required.

According to \citet{shen2007} modulations can be accelerated through fast fourier transforms.
It's implied that the modulation is shift-invariant, in this way the modulation equation is reduced into a convolution equation.
However because of the topology of spherical surface the point spread function (PSF) must satisfy certain requirement so that a convolution can be defined there.

\begin{figure}[htbp]
\centering
\includegraphics[width=0.8\linewidth]{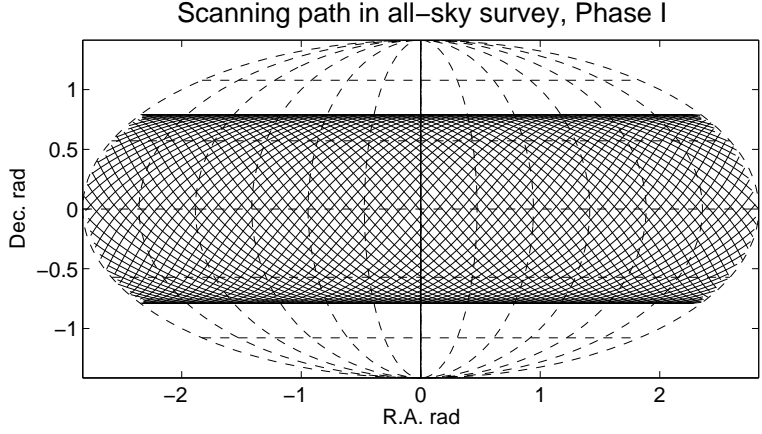}
\caption{Scanning path of the FOV of a detector of HXMT during the first phase of its all-sky survey. We exaggerate the precession rate of the satellite by $20$ times to seperate adjacent scanning circles in this diagram.}
\label{fg-scanning-path}
\end{figure}

The FOVs of HXMT detectors are not circularly symmetric, i.e., contributions to a detector from the sources in its FOV depend upon not only the radial distances from the sources to the center of the FOV, but also the directions.
In addition, during the all-sky survey the FOV of each detector travels on a spherical surface instead of a plane.
As a result, the FOV of each detector around different positions on the celestial sphere are never parallel with each other.
For example, the path of an FOV during the first phase of the all-sky survey is shown in Fig \ref{fg-scanning-path}.
Therefore neither fourier transform nor spherical harmonics can be used to reduce the modulation of kernel function of HXMT observation or data on the celestial sphere.

In the following sections of this article we present a pixelization and tessellation scheme of data defined on spherical surface as well as a method to accelerate the numerical evaluation of modulation on such data thus the DDM for HXMT.

\section{Image reconstruction for HXMT}
\subsection{Pixelization and tessellation of data on spherical surface}
\label{sect-pixel}
As the first step in numerical analysis, pixelization of data on spherical surface is always attractive \citep{tegmark1996,crittenden1998,doroshkevich2005,gorski2005}.
Researches on this problem are driven by applications, e.g., CMB data analysis, since due to the topological nature of a spherical surface,
an theoretically ideal pixelization scheme that would work for all cases does not exist for data defined on such a surface \citep{tegmark1996,gorski2005}.

Image reconstructions for both all-sky survey and observations for selected sky regions can be conquered region by region locally since given a single point of the 
observed data, only objects within the current FOV are relevant while given a single point on the unknown image, only local observed data is relevant.
Therefore all-sky pixelization is not necessary.
We need only a local pixelization scheme adequate for a sky region larger than the FOV of HXMT, which is $5.7^\circ \times 5.7^\circ$.

Virtues of a pixelization scheme in this work include:
\begin{itemize}
\item \textbf{All pixels are of equal size}, which speeds up and simplifies the evaluation of numerical integration, the elementary operation of modulations.

\item Geodesic between two points $p=(i_x,i_y)$ ($i_x$ and $i_y$ are indices) and $q=(i'_x,i'_y)$ as well as azimuth of one point with respect to the other are both 
shift-invariant. That is,
\begin{equation}
d[(i_x,i_y), (i'_x,i'_y)] = d[(i_x+j_x,i_y+j_y), (i'_x+j_x,i'_y+j_y)]\text{,}\;
\forall i_x, i_y, i'_x, j'_y, j_x, j_y\text{,}
\label{eq-distance}
\end{equation}
where $d(p,q)$ is the geodesic distance between $p$ and $q$, i.e., the length of the minor arc on great circle from $p$ to $q$, and,
\begin{equation}
\alpha[(i_x,i_y), (i'_x,i'_y)] = \alpha[(i_x+j_x,i_y+j_y), (i'_x+j_x,i'_y+j_y)]\text{,}\;
\forall i_x, i_y, i'_x, j'_y, j_x, j_y\text{,}
\label{eq-azimuth}
\end{equation}
where $\alpha(p,q)$ is the angle where the geodesic arc from $p$ to $q$ crosses the meridian containing $p$.
Therefore \textbf{all pixels should be of the same shape}.
If the PSF of the telescope is circularly symmetric, i.e., the PSF can be expressed as a function of distance from the center of the FOV, the modulation then 
degenerates to convolution.

\item \textbf{2-D orthogonal pixel indexing}.
Pixels indexed with two indices $i_x$ and $i_y$ in Eq. \ref{eq-distance} and Eq. \ref{eq-azimuth} suggest that the pixel indexing 
consists with a 2-D Cartesian coordinate system.
$I_{i_x, i_y}$ represents $I(i_x s_x, i_y s_y)$, where $s_x$ and $s_y$ are the sampling intervals along $x$-axis and $y$-axis respectively, then $I_{i_x, i_y}$ is the sampled image value on $(i_x, i_y)$.
\end{itemize}

Equidistant cylindrical projection (ECP) method is commonly used in geophysics and climate modeling.
Equidistant pixels on the surface of a cylinder $\boldsymbol{c}$ are projected to the surface of a inscribed sphere $\boldsymbol{s}$ of the cylinder 
$\boldsymbol{c}$ towards their symmetry axis.
Adjacent pixels on the spherical surface are either of the same right ascension (R.A.) or of the same declination (Dec.).
Pixels on the same parallel (or meridian) are spaced on adjacent meridians (or parallels) uniformly.
Pixels with higher latitudes (closer to poles) have smaller sizes and greater distorsions in contrast with pixels with lower latitudes (closer to the equator).

HEALPix by \citet{gorski2005} has recently become a standard structure for spherical data analysis especially for CMB experiments.
Although the equal areas of HEALPix pixels is vital for spherical harmonics transforms, however, the shapes of pixels are different thus the shift-invariance is 
not exactly correct.
Therefore HEALPix is not adapted for HXMT data in this article.
Either ECP or HEALPix method is not adequate for HXMT.

\subsubsection{Quadrilateral projection based pixelization}
\label{sect-pixelization}
A pixelization scheme is designed for HXMT data.
We use the scheme introduced here to build a grid of pixels on a specific sky region.
A \textbf{pixel} on a sphere is an elementary \textbf{area} on the sphere inside its \textbf{boundary} and around its \textbf{center},
the \textbf{boundary} of which on a sphere is defined by $4$ vetices of the pixel and $4$ geodesics of the sphere between each adjacent pairs of them.
The position of the \textbf{center of a pixel} on a sphere is fixed to indicate the area inside the boundary of the pixel.

We start with generating pixels with equal areas and the same shape in a square on a tangent plane $\boldsymbol{p}$ of the unit sphere.
\begin{figure}[htbp]
\pgfdeclarelayer{background}
\pgfdeclarelayer{foreground}
\pgfsetlayers{background,main,foreground}

\centering
\begin{tikzpicture}[scale=2.5,>=stealth]

\begin{pgfonlayer}{background}
\draw (0,0,0) node [anchor=north east] {$O$};
\draw [red,->] (0,0,0) -- (0,2,0) node [anchor=south] {$z$};
\draw [blue] (0,0,0) -- (1,0,0);
\draw [green,->] (0,0,0) -- (0,0,-3) node [anchor=south west] {$y$};
\draw (-1,0,0) node [anchor=east] {$\boldsymbol{s}: x^2+y^2+z^2=1$};
\draw (1,-0.5,-0.5) node [anchor=north west] {$\boldsymbol{p}: x=1$};
\draw [->,violet] (1.5,0.5) -- (1.5cm+1em,0.5) node [black,anchor=west] {{radial projection}};
\draw [->,teal] (1.5,0.5cm+1ex) -- (1.5cm+1em,0.5cm+1ex) node [black,anchor=west] {{parallel projection}};
\draw (1.5cm-0.5em,0.5cm-1ex) rectangle (1.5cm+6em,0.5cm+2ex);
\end{pgfonlayer}

\draw [violet,<-] (0,0) -- (1,-0.25,-0.25);
\draw [violet,<-] (0,0) -- (1,-0.25,0.25);
\draw [violet,<-] (0,0) -- (1,0.25,-0.25);
\draw [violet,<-] (0,0) -- (1,0.25,0.25);

\draw [teal,<-] (0,-0.25,-0.25) -- (1,-0.25,-0.25);
\draw [teal,<-] (0,-0.25,0.25) -- (1,-0.25,0.25);
\draw [teal,<-] (0,0.25,-0.25) -- (1,0.25,-0.25);
\draw [teal,<-] (0,0.25,0.25) -- (1,0.25,0.25);

\shadedraw [shading=ball,opacity=0.66,ball color=brown] (0,0,0) circle (1);

\begin{pgfonlayer}{foreground}
\shadedraw [shading=axis,opacity=0.5,bottom color=orange,top color=white] (1,-0.5,-0.5) -- (1,-0.5,0.5) -- (1,0.5,0.5) -- (1,0.5,-0.5) -- cycle;
\draw (1,0,-0.5) -- (1,0,0.5);
\draw (1,-0.5,0) -- (1,0.5,0);
\draw [blue,->] (1,0,0) -- (2,0,0) node [anchor=west] {$x$};
\end{pgfonlayer}

\end{tikzpicture}
\caption{Plane projection around null position of the surface of a unit sphere. The unit sphere $\boldsymbol{s}$ is centered on $(0,0,0)$, the origin of the $xyz$ 
3-D Cartesian coordinate system. The null position is on $(1,0,0)$. The plane $\boldsymbol{p}$ is parallel to the plane $yOz$. There are 4 pixels on the plane. They 
are either radially or parallelly projected to $\boldsymbol{s}$.}
\label{fg-projection}
\end{figure}
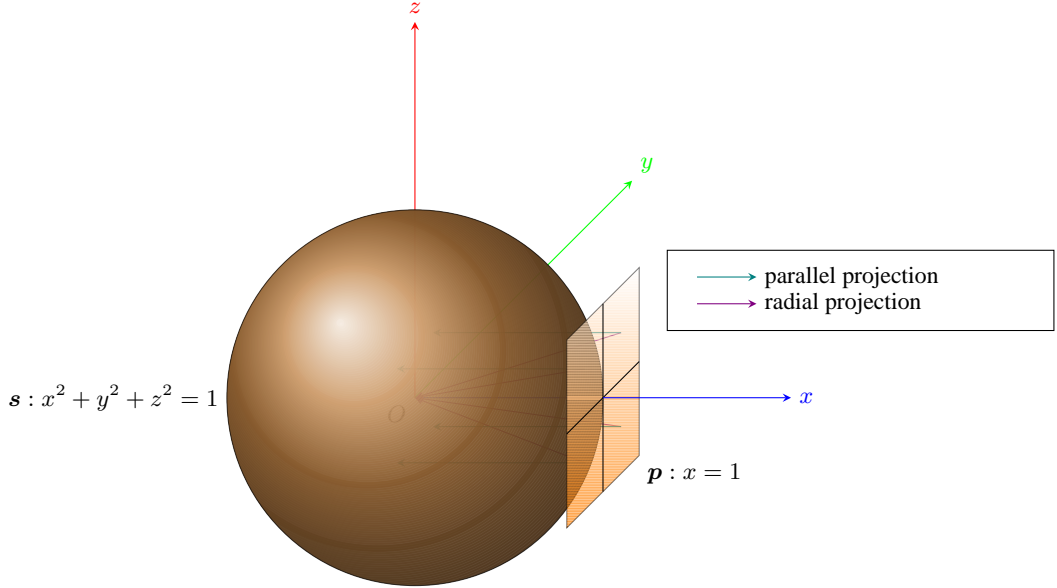
This area is divided into $N_p = N \times N$ square pixels.
Then the centers of all the pixels are projected from the plane $\boldsymbol{p}$ to the surface of the unit sphere $\boldsymbol{s}$ (as shown in Fig. \ref{fg-projection}).
The projection can be either radial or parallel.
In the radial mode, each pixel center is projected towards the center of the sphere $\boldsymbol{s}$, while in the parallel mode it is projected perpendicularly to the plane $\boldsymbol{p}$.

\begin{figure}[htbp]
\centering
%
\pgfdeclarelayer{background}
\pgfdeclarelayer{foreground}
\pgfsetlayers{background,main,foreground}
%
\begin{tikzpicture}
\foreach \x in {0,1,2}
  \foreach \y in {0,1,2} {
  \fill (\x*2cm,\y*2cm) circle (1mm);
  \node[anchor=south west] at (\x*2cm, 4cm-\y*2cm) {$(\y, \x)$};
  }
\node at (2cm, -1cm) {Step 1};
\end{tikzpicture}
%
\hspace{1cm}
\begin{tikzpicture}
\foreach \x in {0,1,2}
  \foreach \y in {0,1,2} {
  \fill (\x*2cm,\y*2cm) circle (1mm);
  \node[anchor=south west] at (\x*2cm, 4cm-\y*2cm) {$(\y, \x)$};
  }

\foreach \x in {0cm,2cm,4cm}
  \draw[dashed] (\x, 0cm) -- (\x, 4cm);

\foreach \y in {0cm,2cm,4cm}
  \draw[dashed] (0cm, \y) -- (4cm, \y);
\node at (2cm, -1cm) {Step 2};
\end{tikzpicture}
\\
%
\vspace{0.5cm}
\begin{tikzpicture}
\foreach \x in {0,1,2}
  \foreach \y in {0,1,2} {
  \fill (\x*2cm,\y*2cm) circle (1mm);
  \node[anchor=south west] at (\x*2cm, 4cm-\y*2cm) {$(\y, \x)$};
  }

\foreach \x in {0cm,2cm,4cm}
  \draw[dashed] (\x, 0cm) -- (\x, 4cm);

\foreach \y in {0cm,2cm,4cm}
  \draw[dashed] (0cm, \y) -- (4cm, \y);

\path plot[mark=x,mark indices={2,4,6,8},mark size=1.0mm] coordinates{(1cm,1cm) (1cm, 2cm) (1cm, 3cm) (2cm, 3cm) (3cm, 3cm) (3cm, 2cm) (3cm, 1cm) (2cm, 1cm) (1cm, 1cm)};
\node at (2cm, -1cm) {Step 3};
\end{tikzpicture}
%
\hspace{1cm}
\begin{tikzpicture}
\begin{pgfonlayer}{background}
\filldraw[draw=black, fill=gray!20!white] plot[mark=x,mark indices={2,4,6,8},mark size=1.0mm] coordinates{(1cm,1cm) (1cm, 2cm) (1cm, 3cm) (2cm, 3cm) (3cm, 3cm) (3cm, 2cm) (3cm, 1cm) (2cm, 1cm) (1cm, 1cm)};
\end{pgfonlayer}
\foreach \x in {0,1,2}
  \foreach \y in {0,1,2} {
  \fill (\x*2cm,\y*2cm) circle (1mm);
  \node[anchor=south west] at (\x*2cm, 4cm-\y*2cm) {$(\y, \x)$};
  }

\foreach \x in {0cm,2cm,4cm}
  \draw[dashed] (\x, 0cm) -- (\x, 4cm);

\foreach \y in {0cm,2cm,4cm}
  \draw[dashed] (0cm, \y) -- (4cm, \y);

\node at (2cm, -1cm) {Step 4};
\end{tikzpicture}
\caption{Steps of definition of a spherical pixel around its center projected from a plane}
\label{fg-pixel-def}
\end{figure}
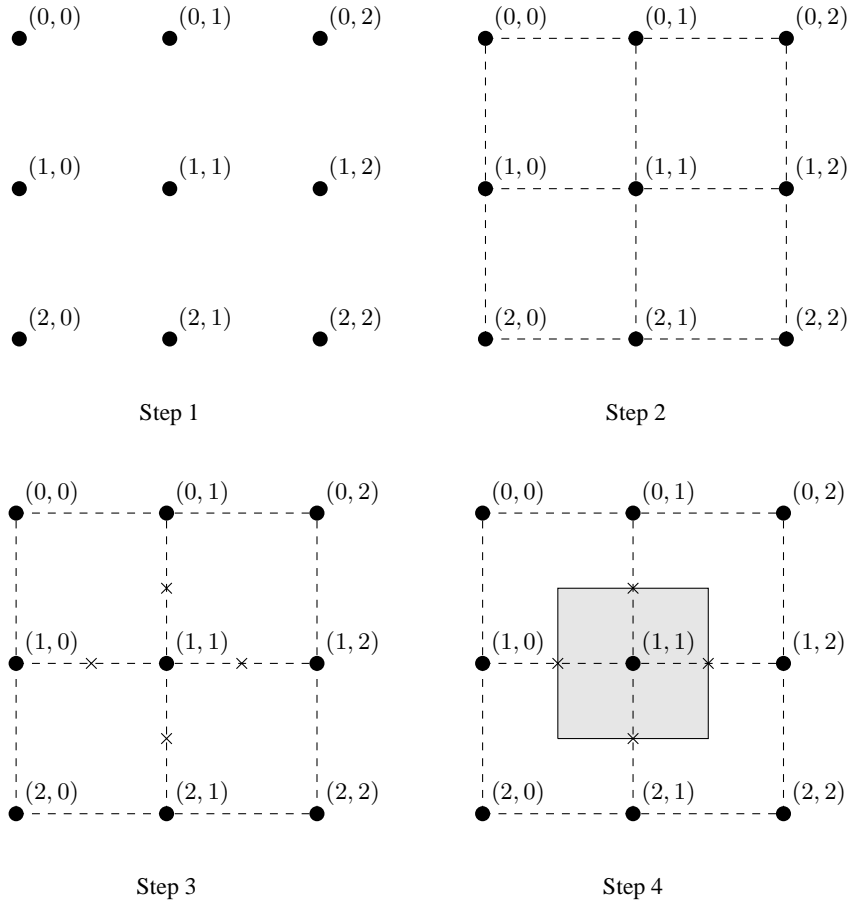

Once all the centers of the pixels are fixed on the sphere (the first step in Fig. \ref{fg-pixel-def}), find the geodesics between each center and its $4$ nearest neighbours (the second step in Fig. \ref{fg-pixel-def}).
Next draw a great-circle arc across the middle point of each geodesic perpendicularly (the third step in Fig. \ref{fg-pixel-def}).
The $4$ points where each perpendicular great-circle arc intersects another two arcs are the vertices and the $4$ perpendicular great-circle arcs make the boundary.
With the center indicating the inside area of the boundary, a pixel is defined on the sphere around each center (the last step in Fig. \ref{fg-pixel-def}).
Pixels on the spherical surface projected far from the square center have greater distorsions in contrast with those projected from the central area of the square.
Obviously the pixel distorsion is independent of the the position of the pixel on the sphere but only depends on its original position in the plane, more 
specifically, the distance from the square center on the plane, therefore we can always use square small enough to make the distorsion negligible.

We use the term \emph{tessella} to refer to a set of pixels on the spherical surface, which are projected from all pixels within a square on a plane.
Hence it suggests that we divide the problem {\textbf{pixelization of a whole spherical surface}} into two problems, {\textbf{pixelization of any small region of 
the spherical surface}} and {\textbf{tessellation of the whole spherical surface using tessellae of pixels}}.
Generally speaking there should not be overlaps or gaps between adjacent tessellae, however, since we will not perform numerical evaluations on different tessellae 
in the same time, overlaps only make evaluations on pixels of the overlaps redundant.

Since all tessellae have the same area and shape, we can generate a set of pixels (i.e. a tessella) around the \emph{null position} of the spherical surface and 
rotate this \emph{initial tessella} to a series of positions to cover the whole sphere.
In this way we can pixelize any small region of a spherical surface by the initial tessella and its rotation expressed by a quaternion (Appendix \ref{app-quat}).

As shown in Fig. \ref{fg-projection}, we define that the null position of the surface of a unit sphere $s$ is $(1,0,0)$ and the plane $\boldsymbol{p}$ is $x=1$.
The center of this square is projected to $(1, 0, 0)$ perpendicularly to $\boldsymbol{p}$.

\begin{figure}[htbp]
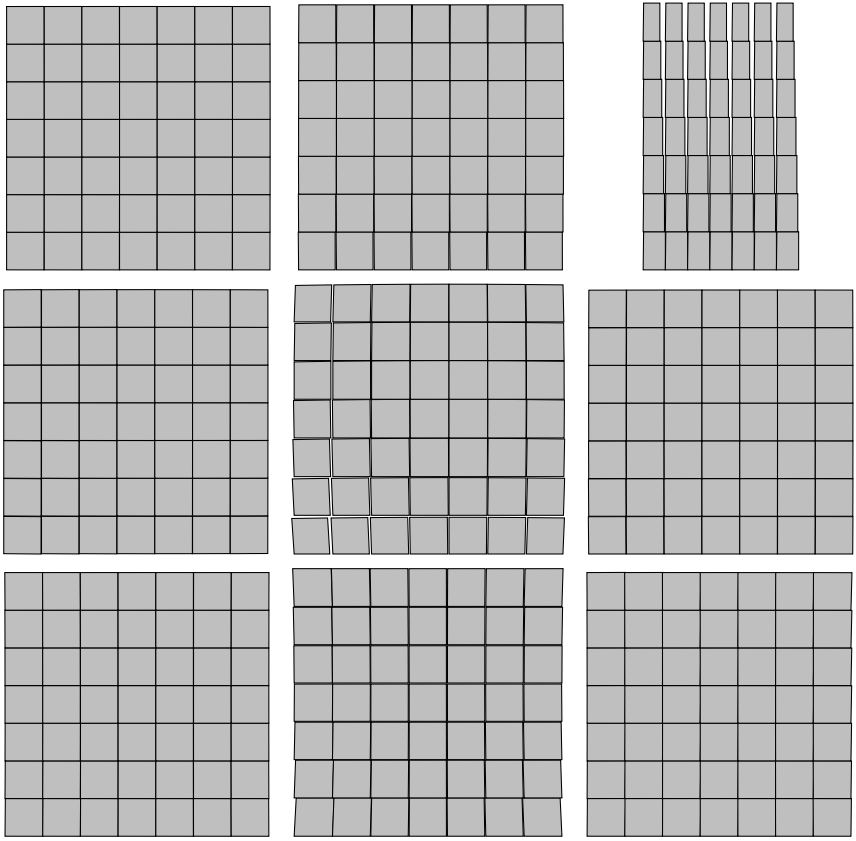

\centering
\hspace{-0.3in}
\begin{tikzpicture}[scale=0.5,>=stealth]
\input{fig0401.tex}
\end{tikzpicture}
\hspace{1.4ex}
\begin{tikzpicture}[scale=0.5,>=stealth]
\input{fig0402.tex}
\end{tikzpicture}
\hspace{0.35in}
\begin{tikzpicture}[scale=0.38,>=stealth]
\input{fig0403.tex}
\end{tikzpicture}\\
\vspace{1ex}
\begin{tikzpicture}[scale=0.5,>=stealth]
\input{fig0404.tex}
\end{tikzpicture}
\hspace{1ex}
\begin{tikzpicture}[scale=0.5,>=stealth]
\input{fig0405.tex}
\end{tikzpicture}
\hspace{1ex}
\begin{tikzpicture}[scale=0.5,>=stealth]
\input{fig0406.tex}
\end{tikzpicture}\\
\vspace{1ex}
\begin{tikzpicture}[scale=0.5,>=stealth]
\input{fig0407.tex}
\end{tikzpicture}
\hspace{1ex}
\begin{tikzpicture}[scale=0.5,>=stealth]
\input{fig0408.tex}
\end{tikzpicture}
\hspace{1ex}
\begin{tikzpicture}[scale=0.5,>=stealth]
\input{fig0409.tex}
\end{tikzpicture}
\caption{\textbf{Top row:} pixel distorsion of equidistance cylindrical projection (left: $11.25^\circ \times 11.25^\circ$, centering at $0^\circ$ latitude;
middle: $30^\circ \times 30^\circ$, centering at $0^\circ$ latitude;
right: $11.25^\circ \times 11.25^\circ$, centering at $60^\circ$ latitude).
\textbf{Middle row:} pixel distorsion of radial plane projection (left: $11.25^\circ \times 11.25^\circ$, centering at $0^\circ$ latitude;
middle: $30^\circ \times 30^\circ$, 
centering at $0^\circ$ latitude;
right: $11.25^\circ \times 11.25^\circ$, centering at $60^\circ$ latitude).
\textbf{Bottom row:} pixel distorsion of parallel plane projection (left: $11.25^\circ \times 11.25^\circ$, centering at $0^\circ$ latitude;
middle: $30^\circ \times 30^\circ$, centering at $0^\circ$ latitude;
right: $11.25^\circ \times 11.25^\circ$, centering at $60^\circ$ latitude).}
\label{fg-distorsion}
\end{figure}

As shown in Fig. \ref{fg-distorsion} the pixel distorsions are negligible in a small region close to the equator, for all the three projection-based pixelization schemes.
The distorsions become significant while the region expends.
Although in low latitude region the distorsion of ECP scheme is suppressed better than those based on plane projection, soon we find that in high latitude regions 
the ECP scheme suffers from severe distorsion, even in a small region. But obviously the distorsions of plane projection schemes are independent of latitudes.

\begin{figure}[htbp]
\centering
\includegraphics[width=0.33\linewidth]{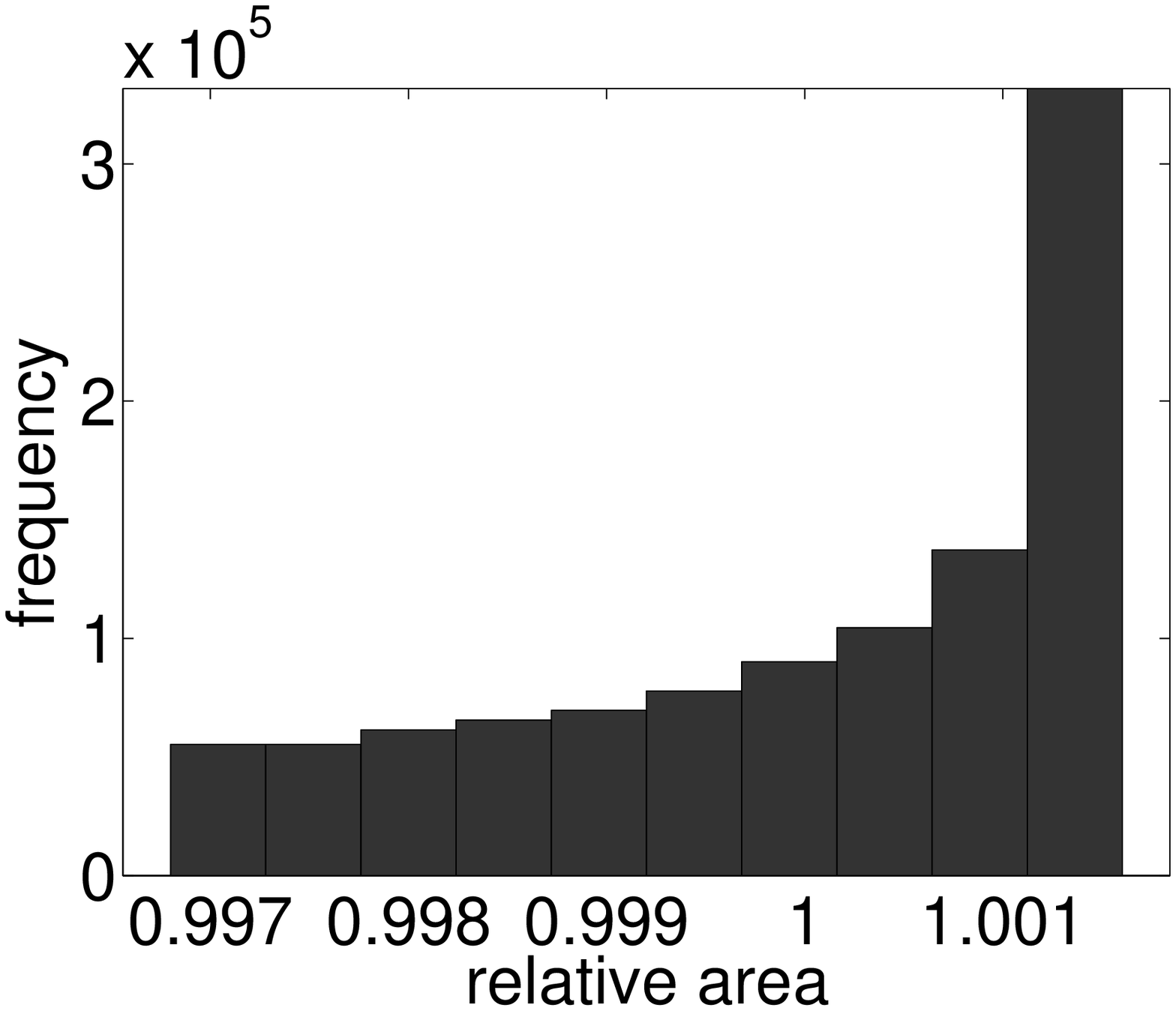}
\includegraphics[width=0.33\linewidth]{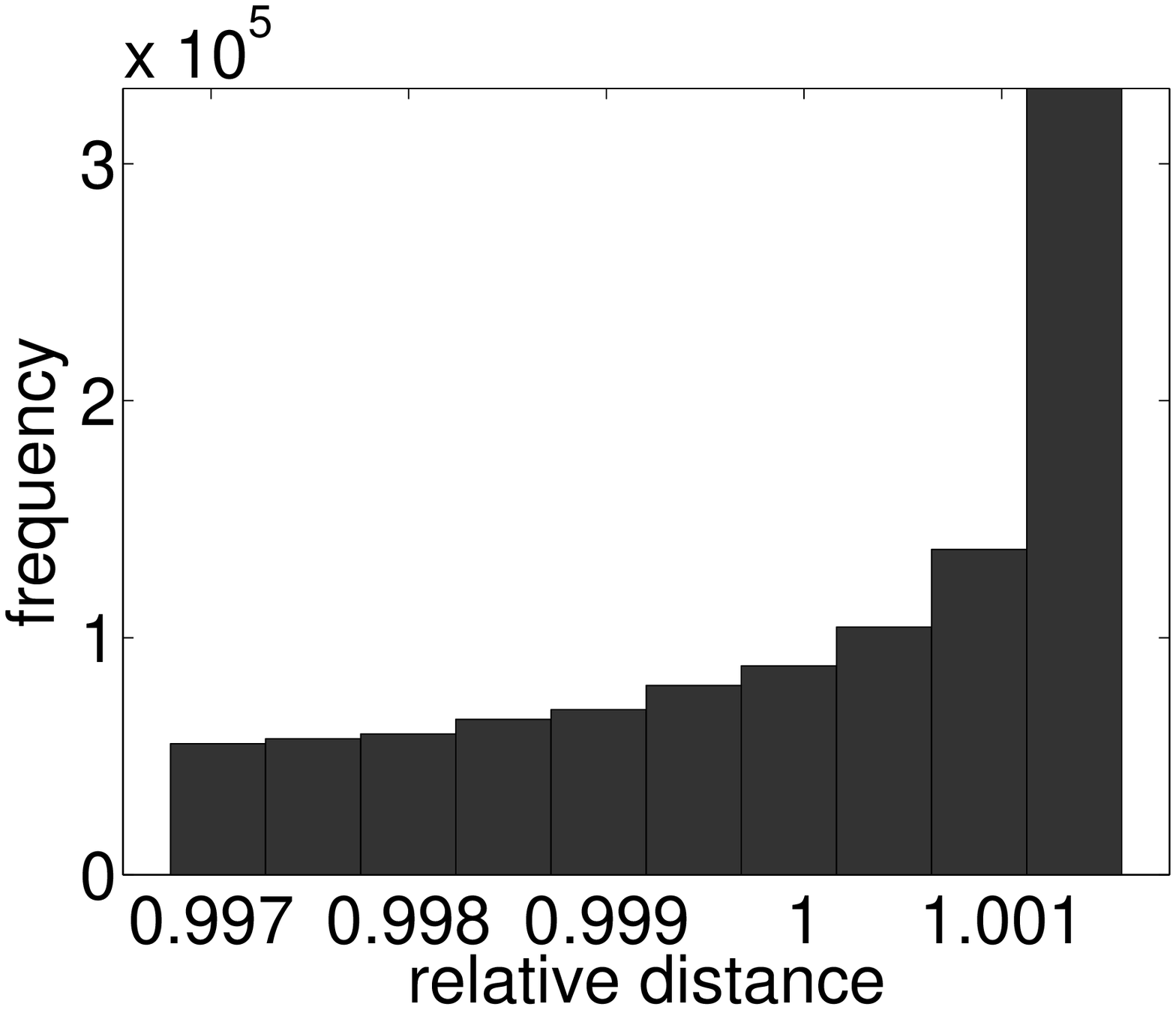}
\includegraphics[width=0.33\linewidth]{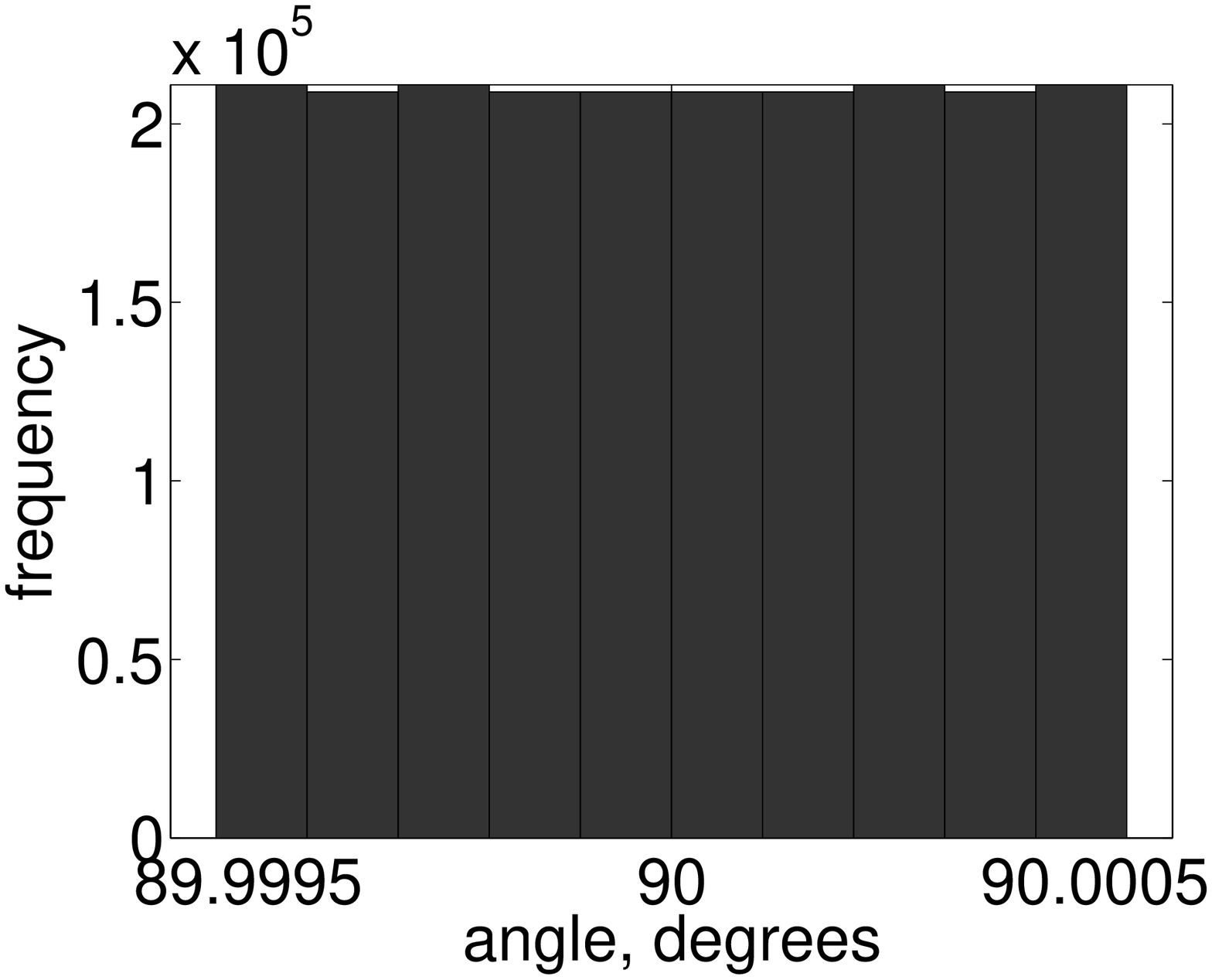}\\
\includegraphics[width=0.33\linewidth]{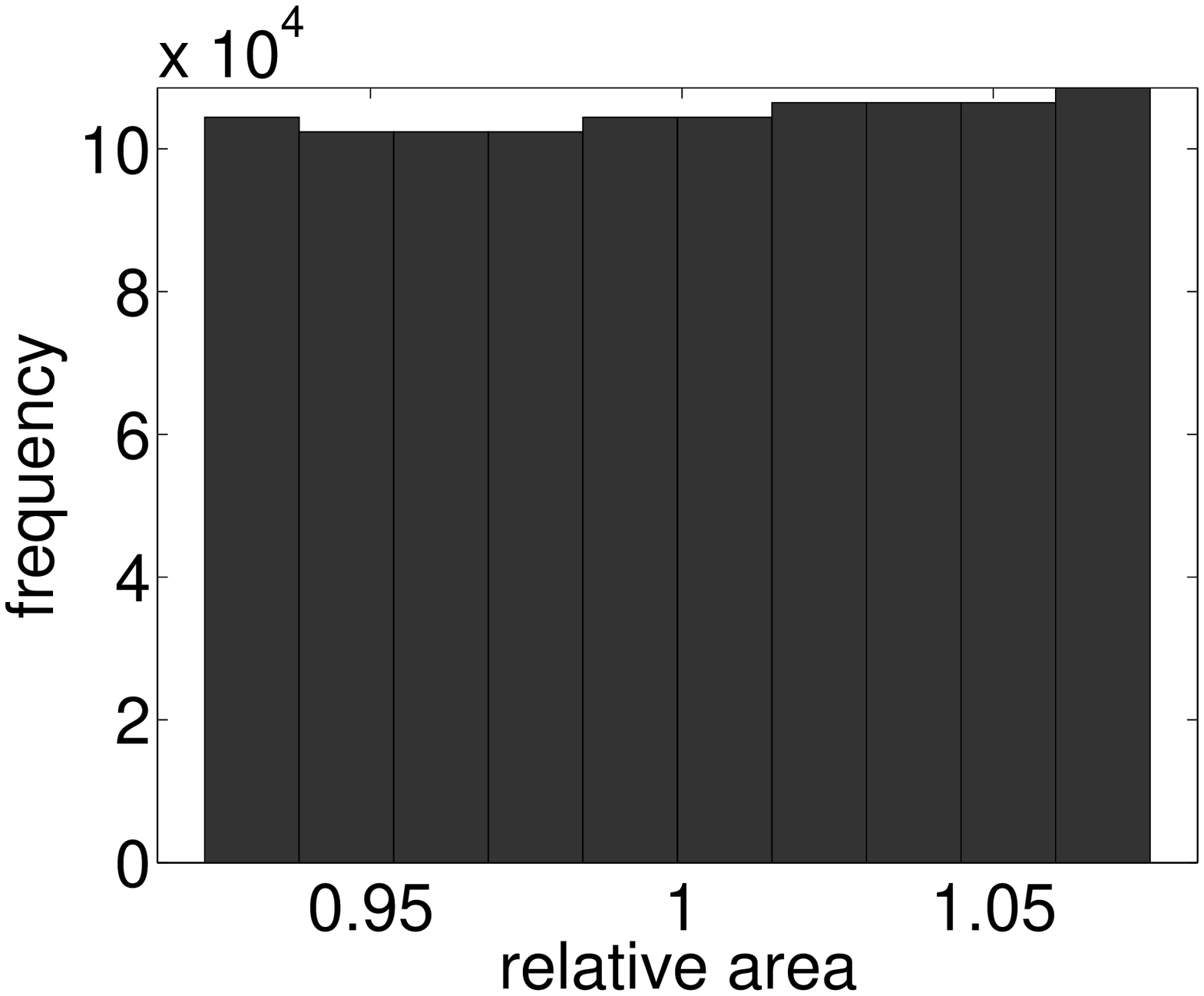}
\includegraphics[width=0.33\linewidth]{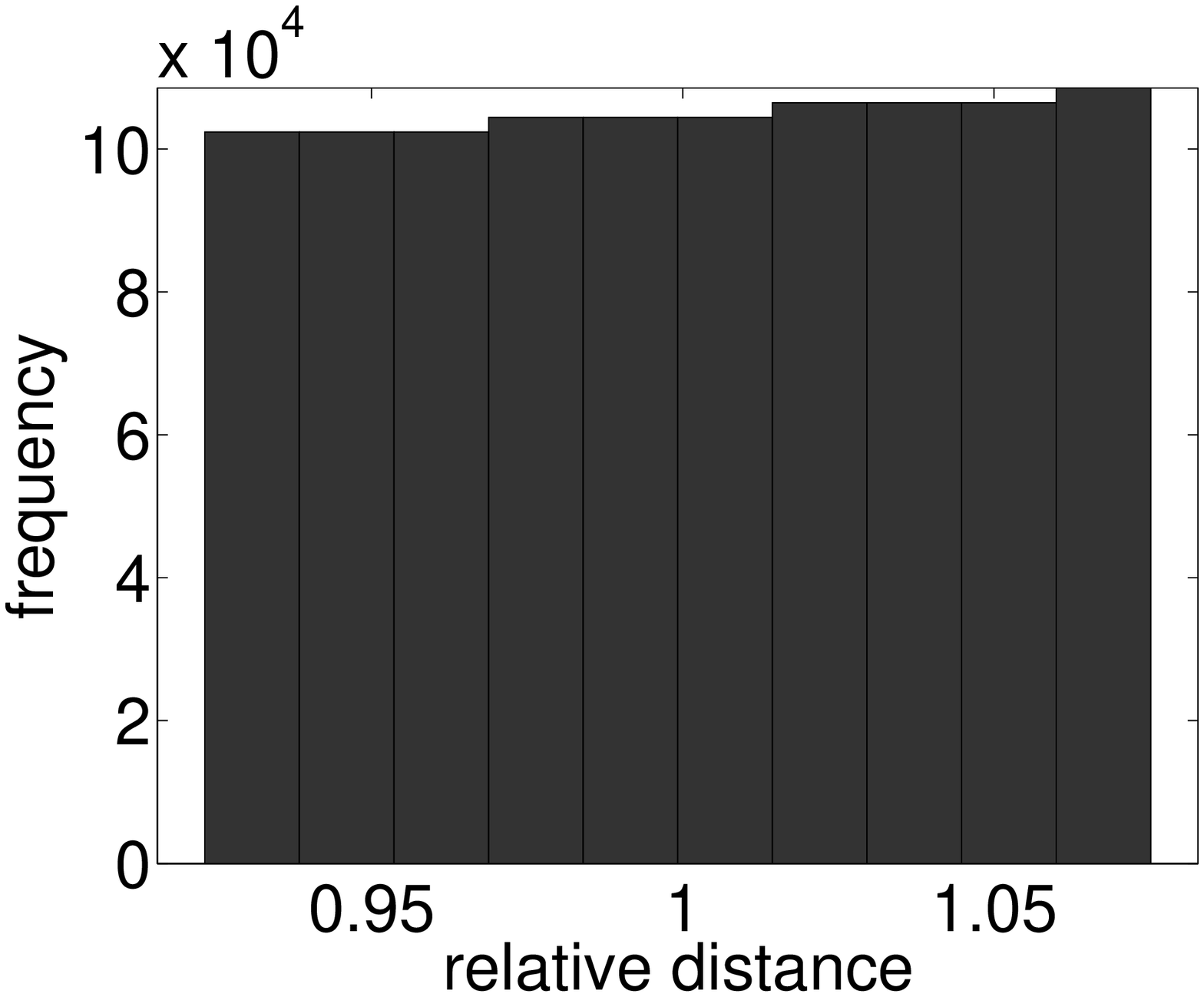}
\includegraphics[width=0.33\linewidth]{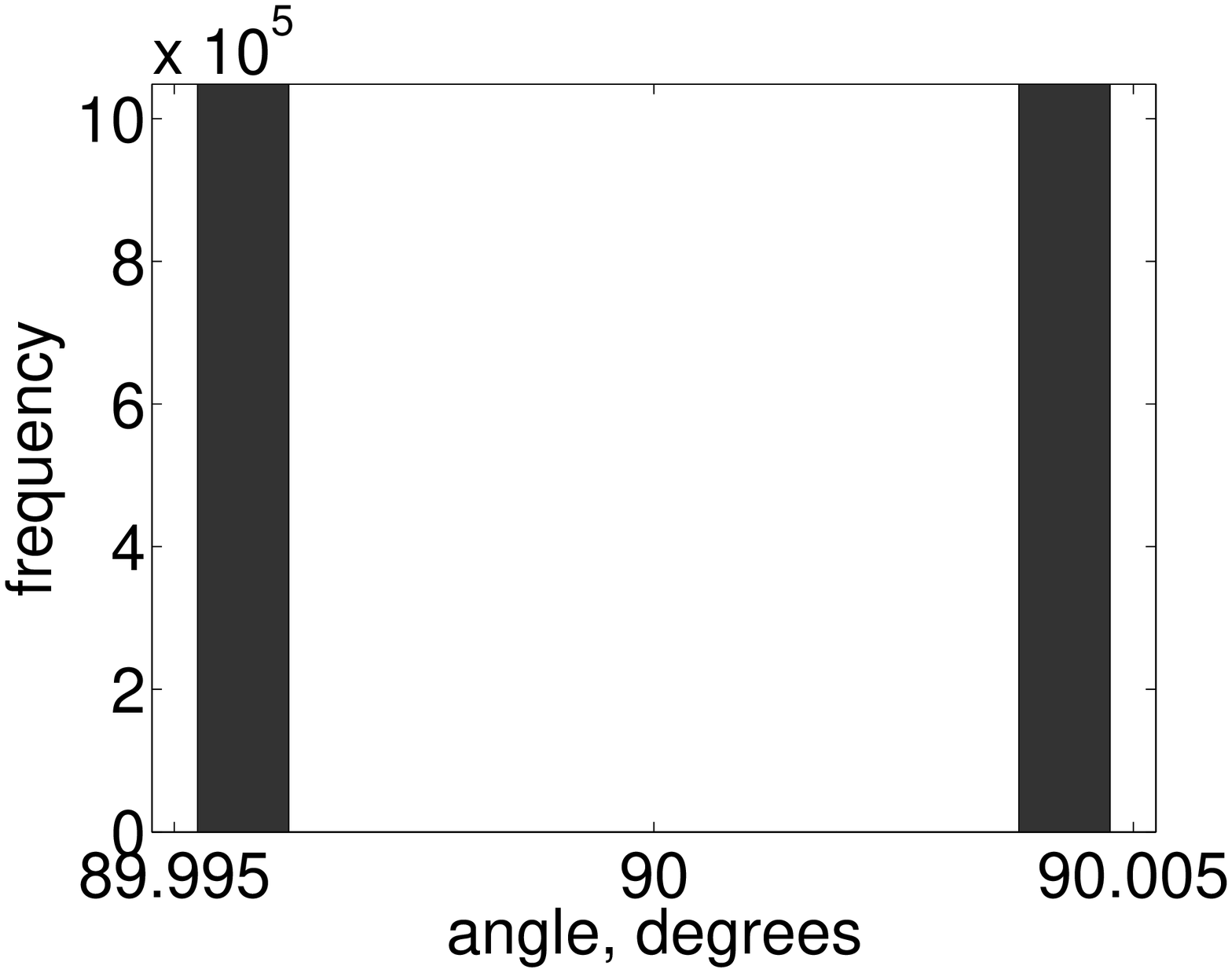}\\
\includegraphics[width=0.33\linewidth]{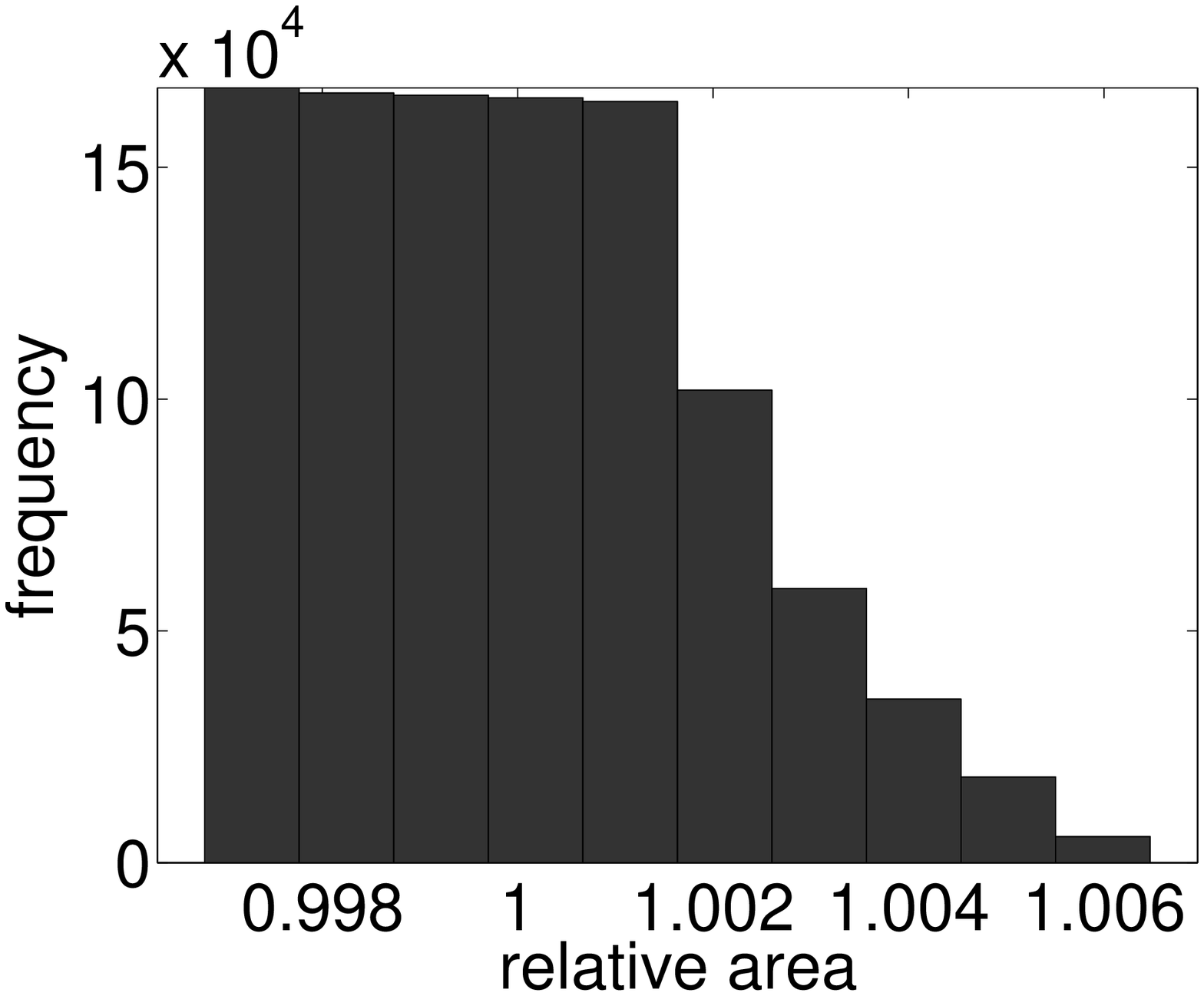}
\includegraphics[width=0.33\linewidth]{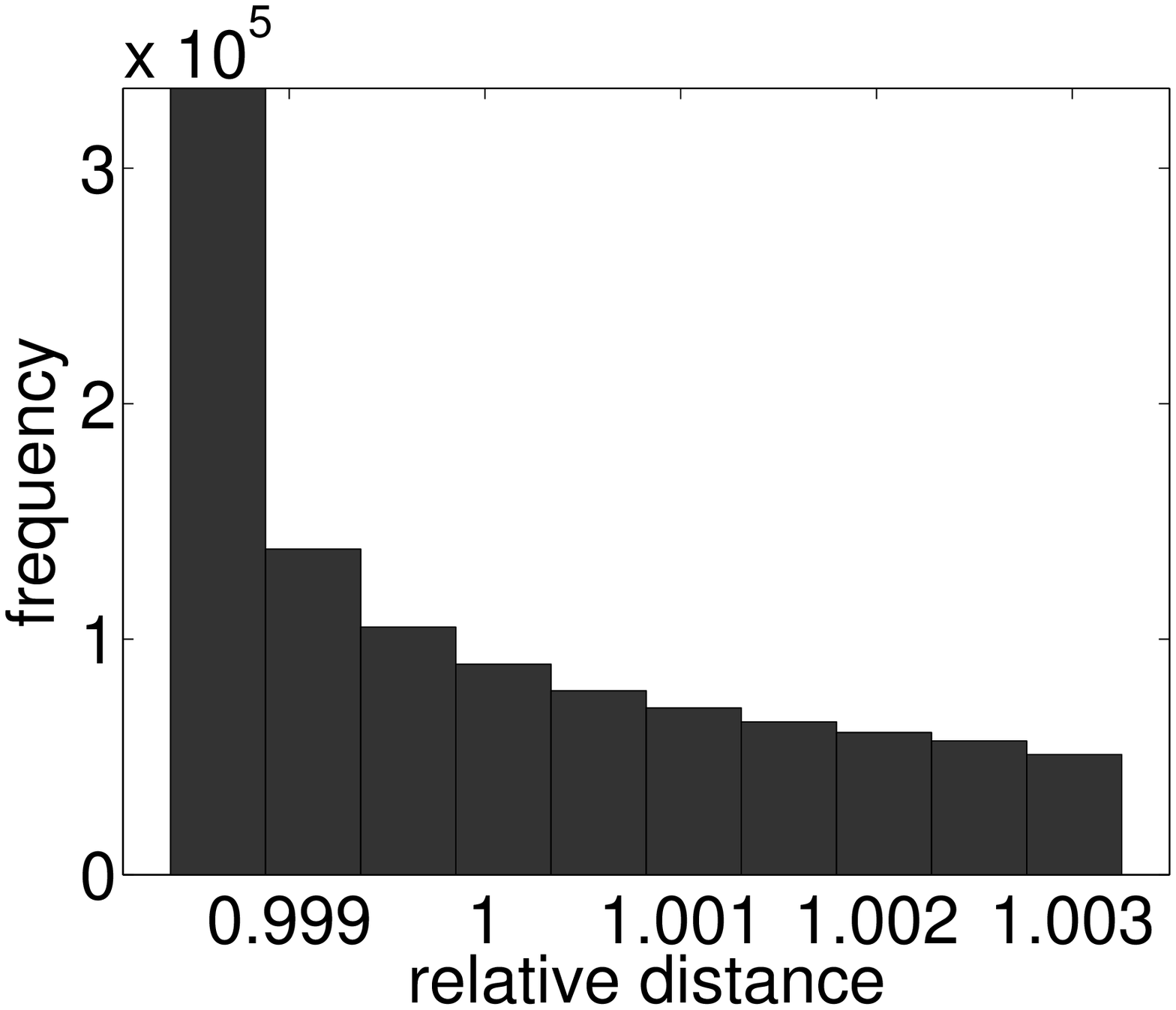}
\includegraphics[width=0.33\linewidth]{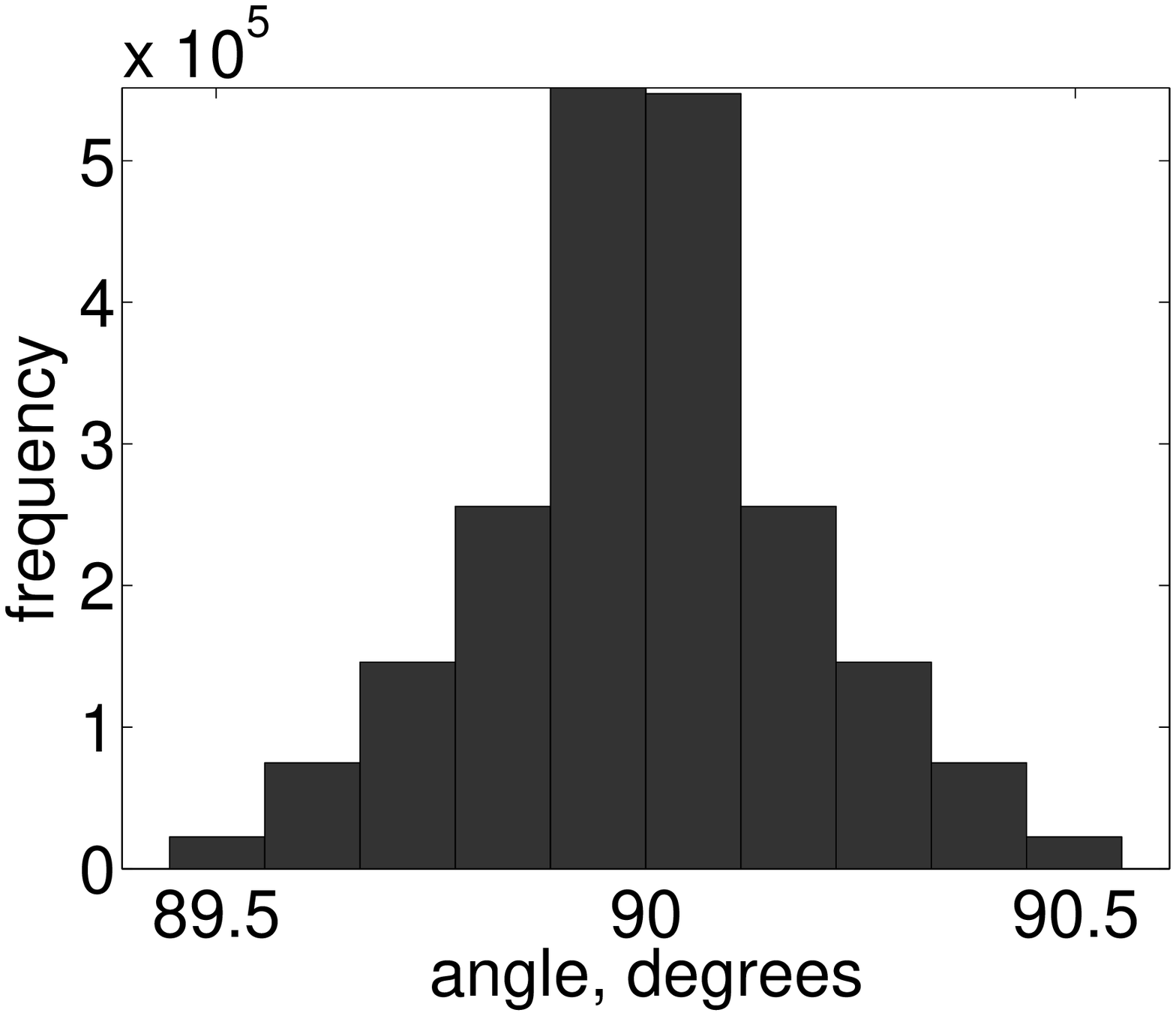}\\
\includegraphics[width=0.33\linewidth]{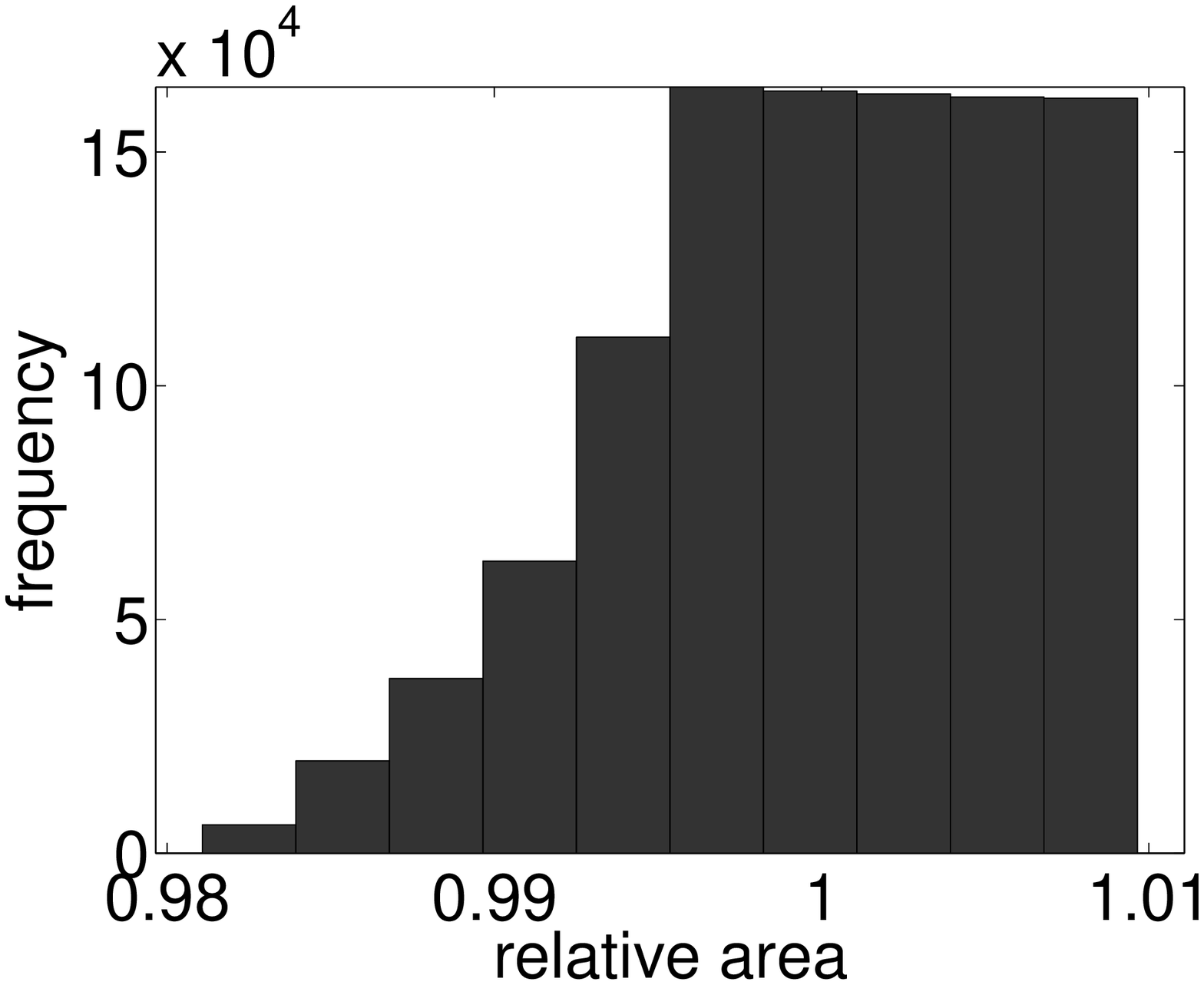}
\includegraphics[width=0.33\linewidth]{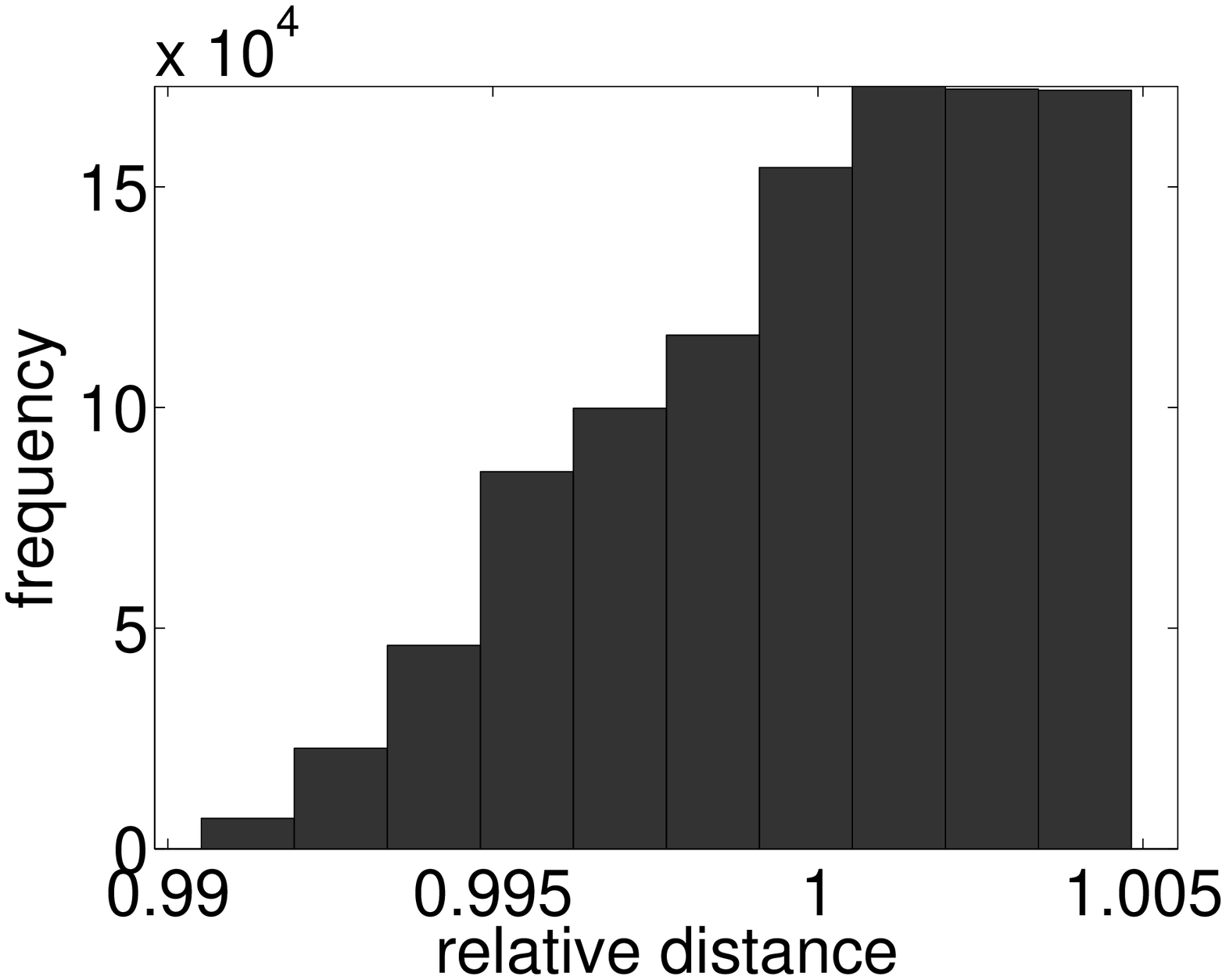}
\includegraphics[width=0.33\linewidth]{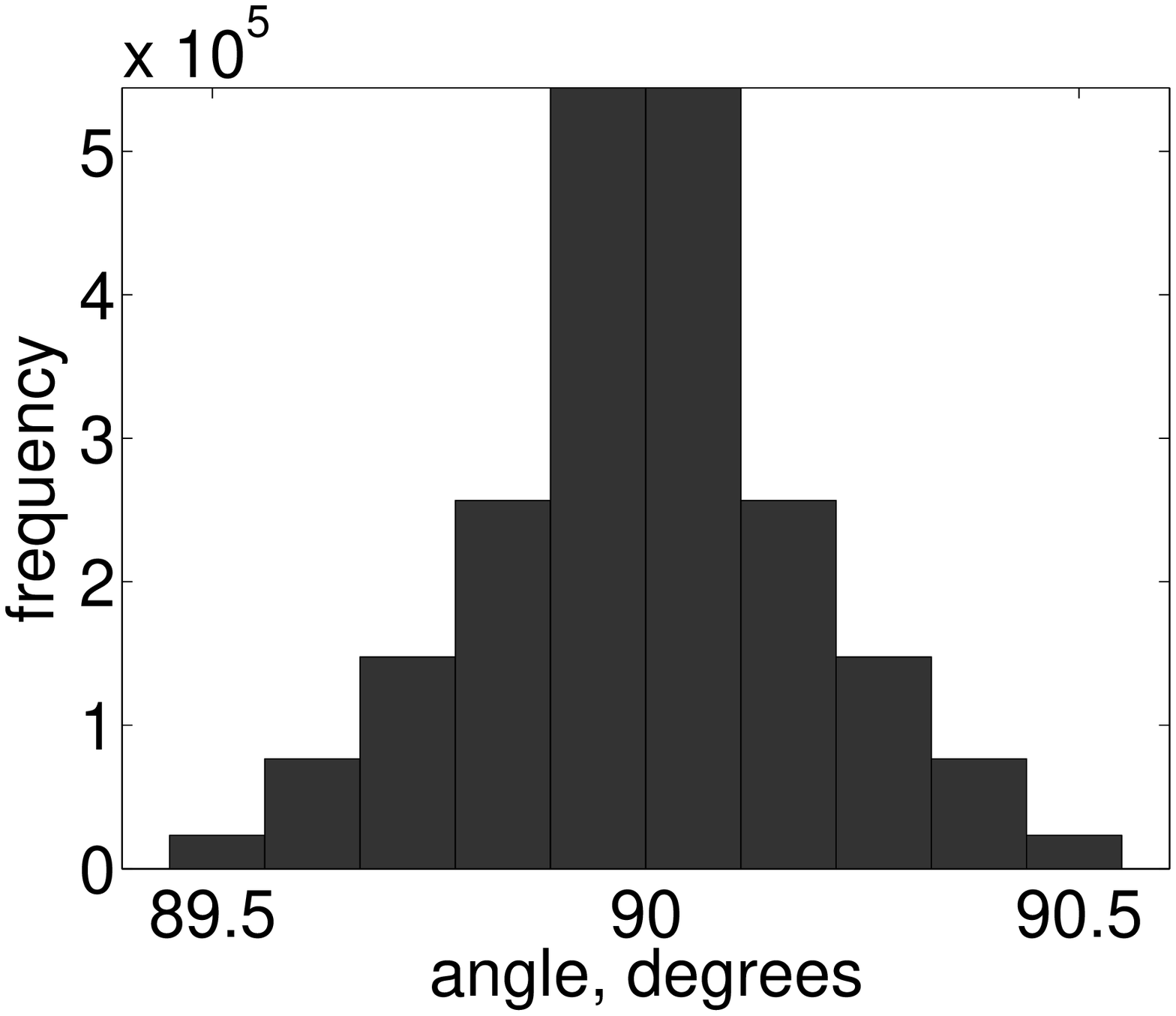}
\caption{Pixel statistics.
$1\,024 \times 1\,024$ pixels are projected to $11.25^\circ \times 11.25^\circ$ spherical surface in different modes.
\textbf{The first row} shows ECP pixels centering at $0^\circ$ latitude;
\textbf{the second row} shows ECP pixels centering at $60^\circ$ latitude;
\textbf{the third row} shows pixels of parallel plane projection;
\textbf{the last row} shows pixels of radial plane projection.
\textbf{The first column} (from left to right) shows relative areas (proportion of the average of areas of all the pixels);
\textbf{the second column} shows relative distances (proportion of the average of distances between adjacent pixels);
\textbf{the last column} shows angles between adjacent sides of the boundary of each pixel.}
\label{fg-pixel-stat}
\end{figure}

See statistics on shapes and areas of pixels on the sphere in Fig. \ref{fg-pixel-stat}.
The relative pixel area distribution is a measure of the uniformity of the pixelization scheme,
which is critical for numerical integration implemented by unweighted summation.
The accuracy of such implementation can be derived from this distribution.
Summation weighted by the relative pixel area can be used to improve the accuracy if necessary.
For a grid of $N \times N$ pixels there are $(N-1) \times (N-1)$ pairs of adjacent pixels.
For each pair we calculate the geodesic distance of the centers of two pixels.
The relative pixel distance distribution is a measure of the uniformity of the sampling intervals.
For each pixel we calculate the angle between the geodesics of its right and down adjacent pixels.
The angle distribution is a measure of the orthogonality of the pixelization scheme.

\subsubsection{Tessellation scheme}
The size of the tessella is $\sigma \times \sigma$, where
\begin{equation}
\sigma = \frac{\pi}{2 M},\;M=1,2,\dots\text{.}
\end{equation}
For the radial projection the size of the square on the tangent plane is $2\tan^{-1}\frac{\sigma}{2} \times 2\tan^{-1}\frac{\sigma}{2}$.
The tessellation of the unit sphere is implemented in 3 steps.

\begin{figure}[htbp]
\centering
\includegraphics[height=2.5in]{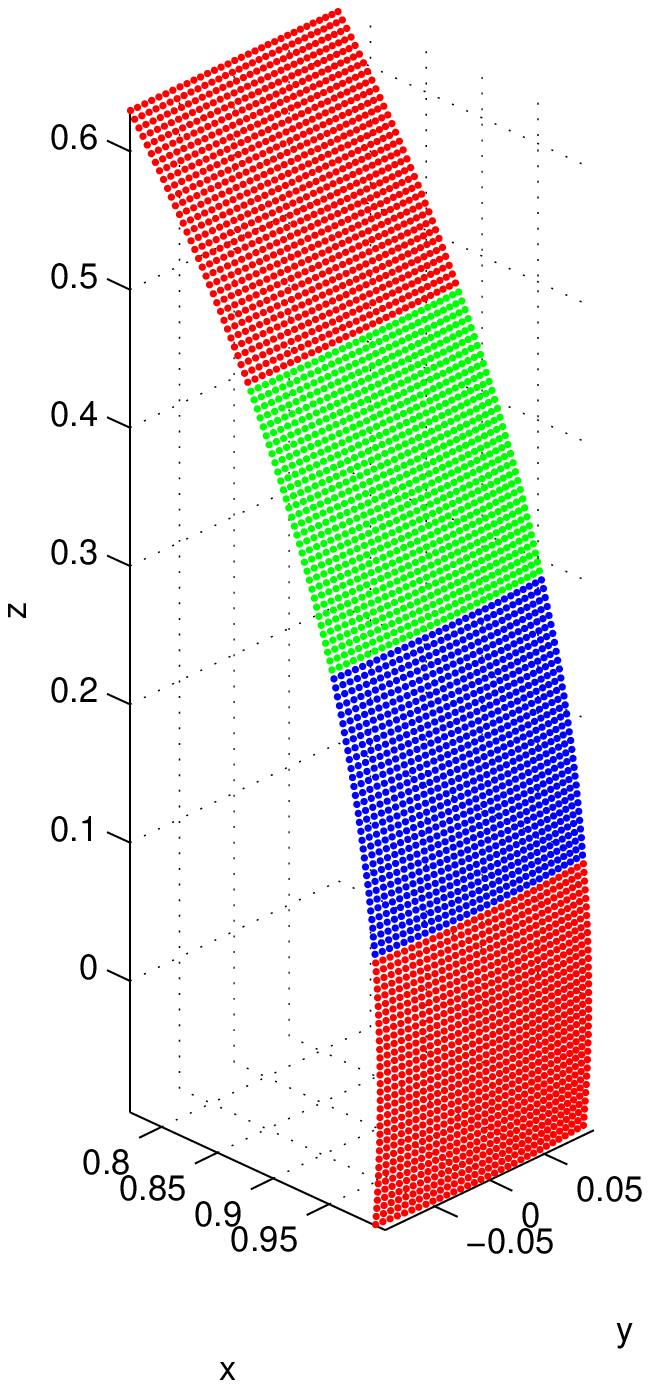}
\includegraphics[height=2.5in]{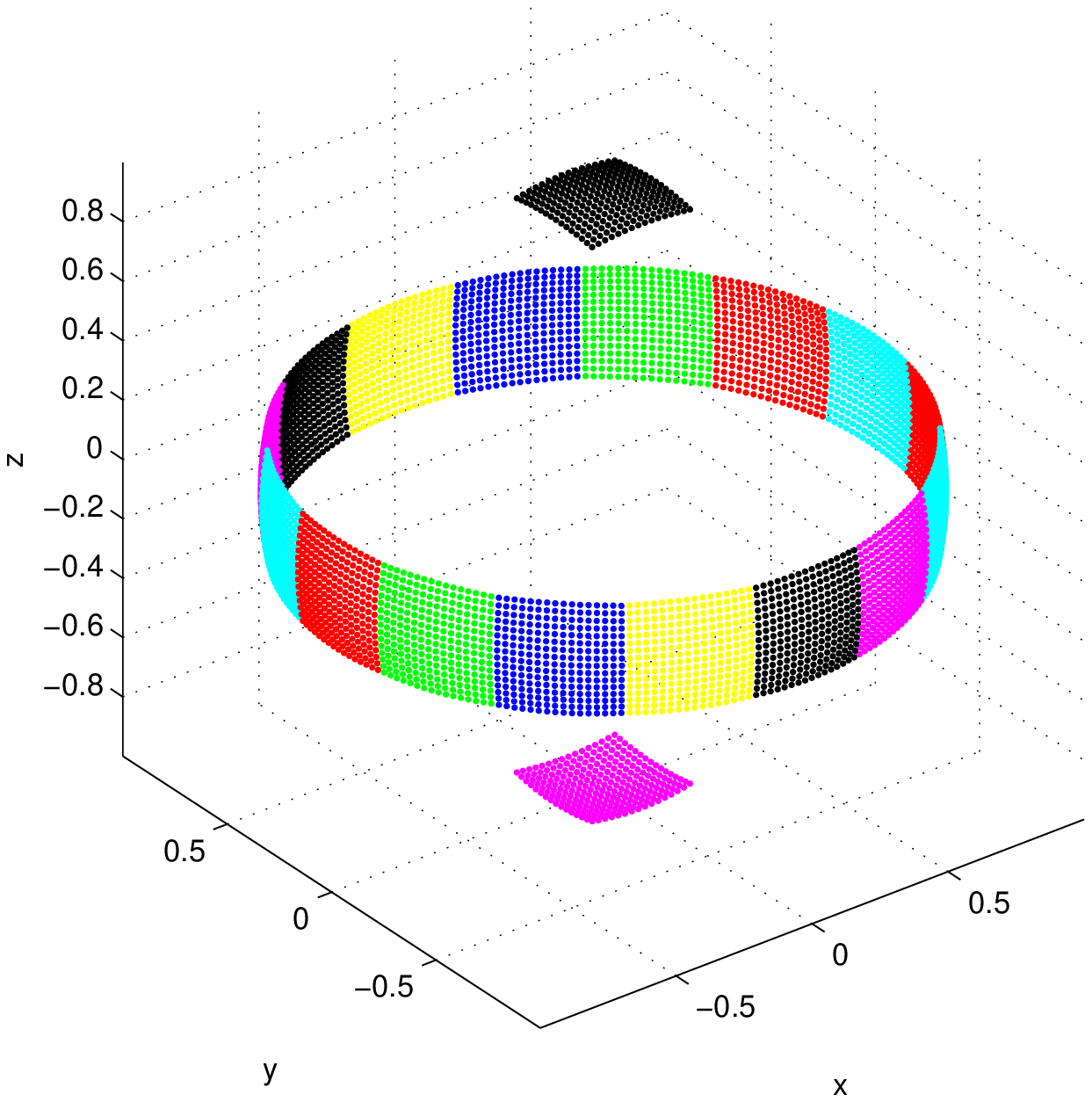}\\
\includegraphics[height=1.5in]{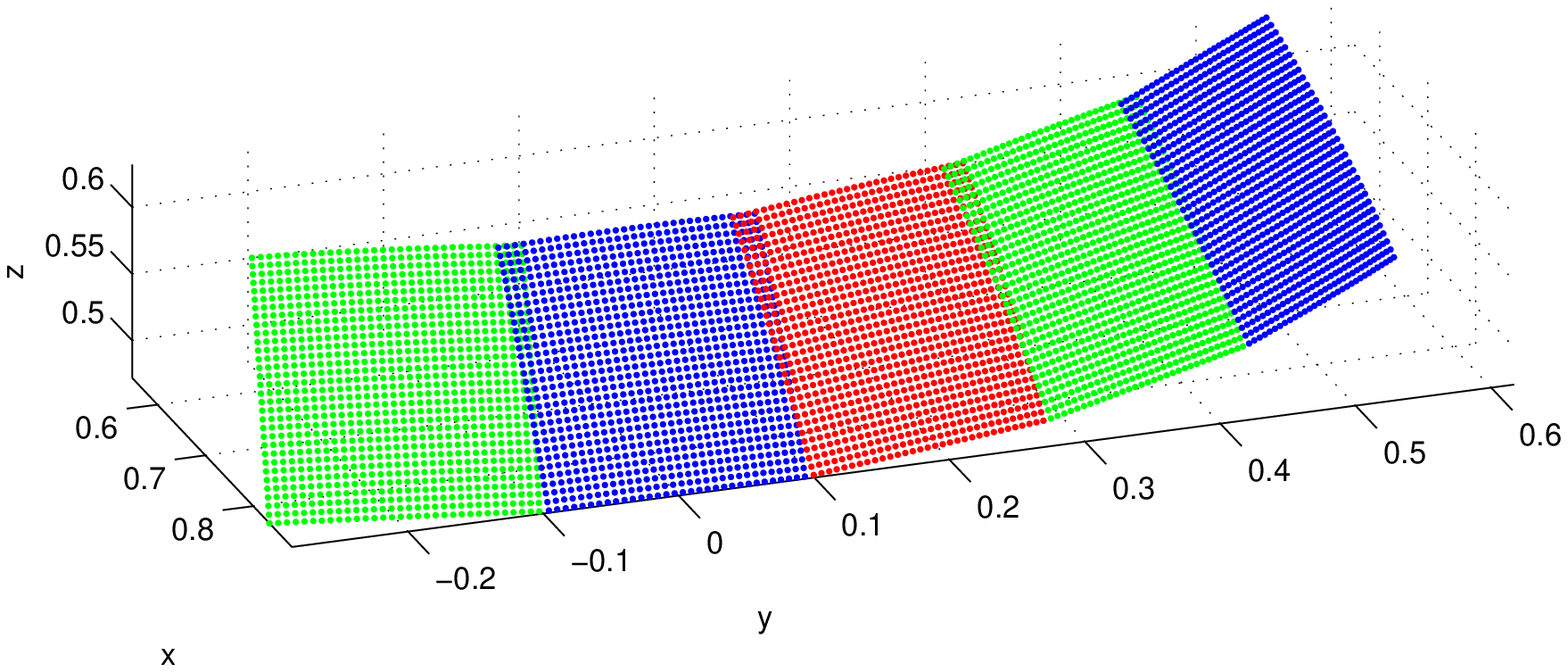}
\caption{Top left: meridianal tessellae on the northern celestial hemisphere;
top right: tessellation of equatorial area and polar area;
bottom: tessellae along a given latitude.}
\label{fg-tessellae}
\end{figure}

The first step is tessellation of the \textbf{prime meridian} of the sphere.
Rotate the initial tessella from the null position towards the north pole as well as the south pole at intervals of $\sigma$, as shown in Fig. 
\ref{fg-tessellae}.
Then the prime meridian is covered by $2M+1$ tessellae.
We call each tessella of those the \emph{meridianal tessella} (Fig. \ref{fg-tessellae}).

The second step is tessellation of the \textbf{equatorial area} of the sphere.
Rotate the initial tessella around $z$-axis also at intervals of $\sigma$ so that the equator of the sphere is covered by $4M$ \emph{equatorial tessellae}, as shown 
in Fig. \ref{fg-tessellae}.

The last step is tessellation of the \textbf{remaining area} of the sphere.
Rotate each meridianal tessella except the initial one and the two covering the polar caps around $z$-axis at intervals of $\frac{\sigma}{\cos(|\theta_w| - 
\frac{\sigma}{2})}$, where $\theta_w$ is the latitude of the center of each tessella, i.e.,
\begin{equation}
\theta_w = \pm\frac{m \pi}{2M}, m=1,\dots,M-1\text{.}
\end{equation}
See the tessellation along a given latitude in Fig. \ref{fg-tessellae}.

\begin{figure}[htbp]
\centering
\includegraphics[width=0.6\linewidth]{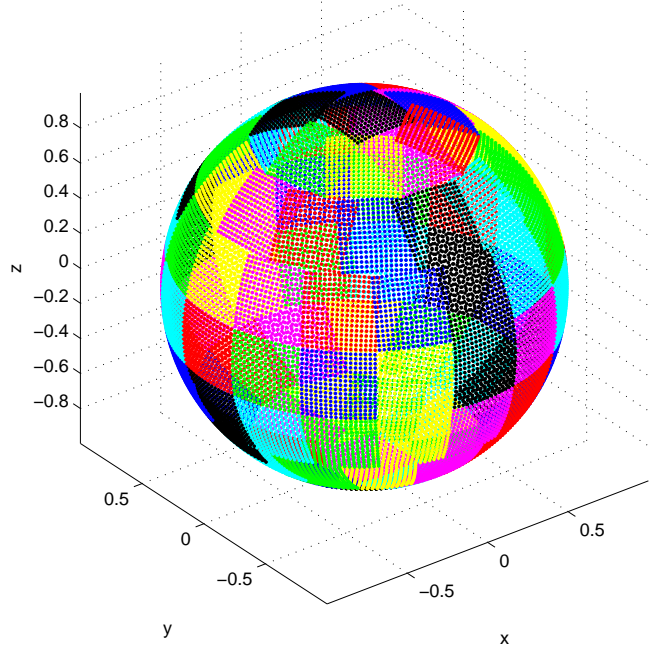}
\caption{Tessellation of the celestial sphere}
\label{fg-global-tessellae}
\end{figure}

Finally the number of tessellae covering the whole sphere is
\begin{equation}
N_t = 4M + 2 + 2\times\sum_{m=1}^{M-1}\bigg\lceil \pi / \arctan\bigr(\frac{\tan\frac{\pi}{4M}}{\cos\frac{m\pi}{2M} + \sin\frac{m\pi}{2M}\tan\frac{\pi}{4M}}\bigl) 
\bigg\rceil\text{,}
\end{equation}
where $\lceil \cdots \rceil$ is the ceiling operator denoting the smallest integer no less than the inside number.
See the tessellation of the celestial sphere in Fig. \ref{fg-global-tessellae}.

The quaternion representing the rotation upon each tessella is
\begin{equation}
\boldsymbol{q}_{\phi_w, \theta_w} = 
\cos\frac{\theta_w}{2} \cos\frac{\phi_w}{2} + 
\sin\frac{\theta_w}{2} \sin\frac{\phi_w}{2} \mathrm{i} -
\sin\frac{\theta_w}{2} \cos\frac{\phi_w}{2} \mathrm{j} +
\cos\frac{\theta_w}{2} \sin\frac{\phi_w}{2} \mathrm{k}\text{,}
\end{equation}
where $\phi_w$ and $\theta_w$ represent the position of the center of the tessella on the sphere, in longitude and latitude respectively.
The corresponding rotation matrix is
\begin{equation}
\boldsymbol{R}_{\phi_w,\theta_w} = \begin{pmatrix}
\cos\theta_w \cos\phi_w & -\sin\phi_w & -\sin\theta_w \cos\phi_w \\
\cos\theta_w \sin\phi_w &  \cos\phi_w & -\sin\theta_w \sin\phi_w \\
\sin\theta_w            &  0          &  \cos\theta_w
\end{pmatrix}\text{.}
\end{equation}
Given a pixel $\boldsymbol{r}_{i,j}$ on the initial tessella, where $i$ and $j$ are the indices of each pixel, the corresponding pixel on the tessella around $(\phi_w, \theta_w)$ is then $\boldsymbol{R}_{\phi_w, \theta_w}\boldsymbol{r}_{i,j}$.

\subsubsection{Pixelization of observed data}
The original observed data from each detector module is recorded as a TOA (time of arrival) sequence of x-ray photons.
It is required to convert the original observed data from time series into pixel-wise format, i.e., pixelization of the observed data, for following processes.
Alongside with the scientific data there is also engineering data being recorded such as the attitude of the spacecraft etc.
By interpolations the attitudes of the spacecraft can be determined at any given point during the observation time, i.e.,
for each detected photon the attitude of the spacecraft can be assigned.
From the attitude of the spacecraft we can calculate the position on the celestial sphere where each collimator is pointing.
For example, in the equatorial coordinate system, given a specific detector module, each detected photon from this module comes with not only its TOA but also the R.A. and Dec. of a position on the celestial sphere the collimator is pointing at, i.e., the projection on the celestial sphere along the optical axis of the collimator.
The projection on the celestial sphere is then called the position of the detected photon.

For each photon first we find the tessallae that cover its position on the celestial sphere.
Then for each tessella that covers its position, we find the pixel that takes up this position.
Because the boundary of each tessella as well as for each pixel is defined, the problem of finding whether a specific point on the celestial sphere lies inside or outside a tessella or a pixel is classified as a \textbf{point-in-square-on-sphere problem}. See in Appendix \ref{app-pisos} our approach to solve this problem.

Once the position of each detected photon is mapped to pixels and tessellae we defined on the celestial sphere, the \textbf{total number of detected photons} can be calculated for each pixel.
Taking the time interval between the time of arrival of a photon and the next one on the same detector as the exposure time of the former photon, the exposure time of all the photons mapped to each pixel together yields the \textbf{total exposure time} on this pixel.
Divide the total number of photons by the total exposure time on each pixel, and we have the \textbf{counting rate} for each pixel, i.e., the observed data in pixel-wise format.

\subsection{Acceleration of modulation}
\subsubsection{Status parameters of the collimated detector}
Evaluating numerical modulation is ubiquitous in both simulation of observed data and image reconstruction with direct demodulation method.
In HXMT observation the observed data is a function of the given collimator status, which includes the \emph{collimator identifier} $c$ and its orientation.

To describe the orientation of a given collimator, we use a unit vector along its optical axis as well as another unit vector along a fixed axis perpendicular to its optical axis (e.g., long edge of one of its slice).
We call the unit vector along its optical axis and the later one \textbf{pointing vector} and \textbf{position vector} (imitating the position angle parameter of a telescope)  respectively.
Then we define a null status of a given collimator where its pointing vector points at the null position of the current coordinate system (along $x$-axis) and its position vector points at $(0, 0, 1)$ (along $z$-axis).
We use three status parameters $\phi$, $\theta$ and $\psi$ to represent the current status of the given collimator.
First rotate the given collimator around $z$-axis from its null position by $\phi$ then incline it from the $xOy$ plane (e.g. the equator) by $\theta$, and finally rotate it around its own optical axis by $\psi$ i.e., the position angle.
In this way the collimator can be rotated to any possible orientation.

We use a quaternion $\boldsymbol{q}$ to formulate the rotation parameterized by $\phi$, $\theta$ and $\psi$ on the pointing vector and position vector of a given collimator as
\begin{align}
\boldsymbol{q}_1 & = \cos\frac{\phi}{2} + \sin\frac{\phi}{2}\mathrm{k}\text{,}\\
\boldsymbol{q}_2 & = \cos\frac{\theta}{2} + \sin\phi \sin\frac{\theta}{2}\mathrm{i} - \cos\phi \sin\frac{\theta}{2}\mathrm{j}\text{,}\\
\boldsymbol{q}_3 & = \cos\frac{\psi}{2} + \cos\phi \cos\theta \sin\frac{\psi}{2}\mathrm{i} + \sin\phi \cos\theta \sin\frac{\psi}{2}\mathrm{j} + \sin\theta \sin\frac{\psi}{2}\mathrm{k}\text{,}\\
\boldsymbol{q}   & = \boldsymbol{q}_3 \boldsymbol{q}_2 \boldsymbol{q}_1\text{,}
\label{eq-etoq}
\end{align}
where $\boldsymbol{q}_1$, $\boldsymbol{q}_2$ and $\boldsymbol{q}_3$ are auxiliary quaternions, and $\boldsymbol{q}$ rotates the collimator from its null status $\Bigr(\begin{smallmatrix}
0\\
0\\
0
\end{smallmatrix}\Bigl)$ to $\Bigr(\begin{smallmatrix}
\phi\\
\theta\\
\psi
\end{smallmatrix}\Bigl)$.
The corresponding rotation matrix is
\begin{equation}
\boldsymbol{R}_{\phi,\theta,\psi}=\begin{pmatrix}
\cos\theta \cos\phi & - \sin\phi \cos\psi - \sin\theta \cos\phi \sin\psi &   \sin\phi \sin\psi - \sin\theta \cos\phi \cos\psi \\
\cos\theta \sin\phi &   \cos\phi \cos\psi - \sin\theta \sin\phi \sin\psi & - \cos\phi \sin\psi - \sin\theta \sin\phi \cos\psi \\
\sin\theta          &                       \cos\theta          \sin\psi &                       \cos\theta          \cos\psi
\end{pmatrix}\text{.}
\end{equation}

Given the status parameters $\phi$, $\theta$ and $\psi$, the corresponding pointing vector and position vector are
\begin{equation}
\boldsymbol{p}_{\phi,\theta,\psi} = \boldsymbol{R}_{\phi,\theta,\psi}\begin{pmatrix}
1\\
0\\
0
\end{pmatrix} = 
\begin{pmatrix}
\cos\theta \cos\phi\\
\cos\theta \sin\phi\\
\sin\theta
\end{pmatrix}\text{,}
\label{eq-pointing}
\end{equation}
and
\begin{equation}
\boldsymbol{a}_{\phi,\theta,\psi} = \boldsymbol{R}_{\phi,\theta,\psi}\begin{pmatrix}
0\\
0\\
1
\end{pmatrix} = 
\begin{pmatrix}
  \sin\phi \sin\psi - \sin\theta \cos\phi \cos\psi\\
- \cos\phi \sin\psi - \sin\theta \sin\phi \cos\psi\\
                      \cos\theta          \cos\psi
\end{pmatrix}\text{,}
\label{eq-azimuthal}
\end{equation}
respectively.

We call the vector $\Bigr(\begin{smallmatrix}
\phi\\
\theta\\
\psi
\end{smallmatrix}\Bigl)$ the status vector of a given collimator, where $\phi$, $\theta$ are the two pointing parameters and $\psi$ is the position angle parameter.
The status parameters are coordinate-system-dependent.
Null status vector $\Bigr(\begin{smallmatrix}
0\\
0\\
0
\end{smallmatrix}\Bigl)$ in different coordinate systems stands for different orientations of the same collimator.

For example, if the equatorial coordinate system is adopted, $\psi=0$ means its position vector and pointing vector are on the same longitude of the celestial sphere, i.e., the R.A. of both its position vector and pointing vector are equal to each other.
The angle between the two vectors is always $\frac{\pi}{2}$ by definition.
So in this case the Dec. of its position vector differs by $90^\circ$ from the Dec. of its pointing vector.

\subsubsection{PSF and numerical evaluation of the modulation kernel function}
In the equatorial coordinate system an image is a function of positions on the celestial sphere denoted by R.A. and Dec.

Given the observed data $d(\phi,\theta,\psi,c)$ and the image $f(\phi,\theta)$, the modulation in Eq. \ref{eq-modulate} yields
\begin{equation}
d(\phi,\theta,\psi,c) = \int_{\Omega} p(\phi,\theta,\psi,c,\phi',\theta') f(\phi',\theta') \mathrm{d}\Omega\text{,}
\end{equation}
where $p(\phi,\theta,\psi,c,\phi',\theta')$ is the modulation kernel function represents the response of collimator $c$ with status $\Bigr(\begin{smallmatrix}\phi\\ \theta\\ \psi\end{smallmatrix}\Bigl)$ to a unit point object at position $\Bigr(\begin{smallmatrix}\cos\theta' \cos\phi' \\
\cos\theta' \sin\phi' \\
\sin\theta' \end{smallmatrix}\Bigl)$, and $\Omega$ respresents the solid angle.

\begin{figure}[htbp]
\centering
\includegraphics[width=0.3\linewidth]{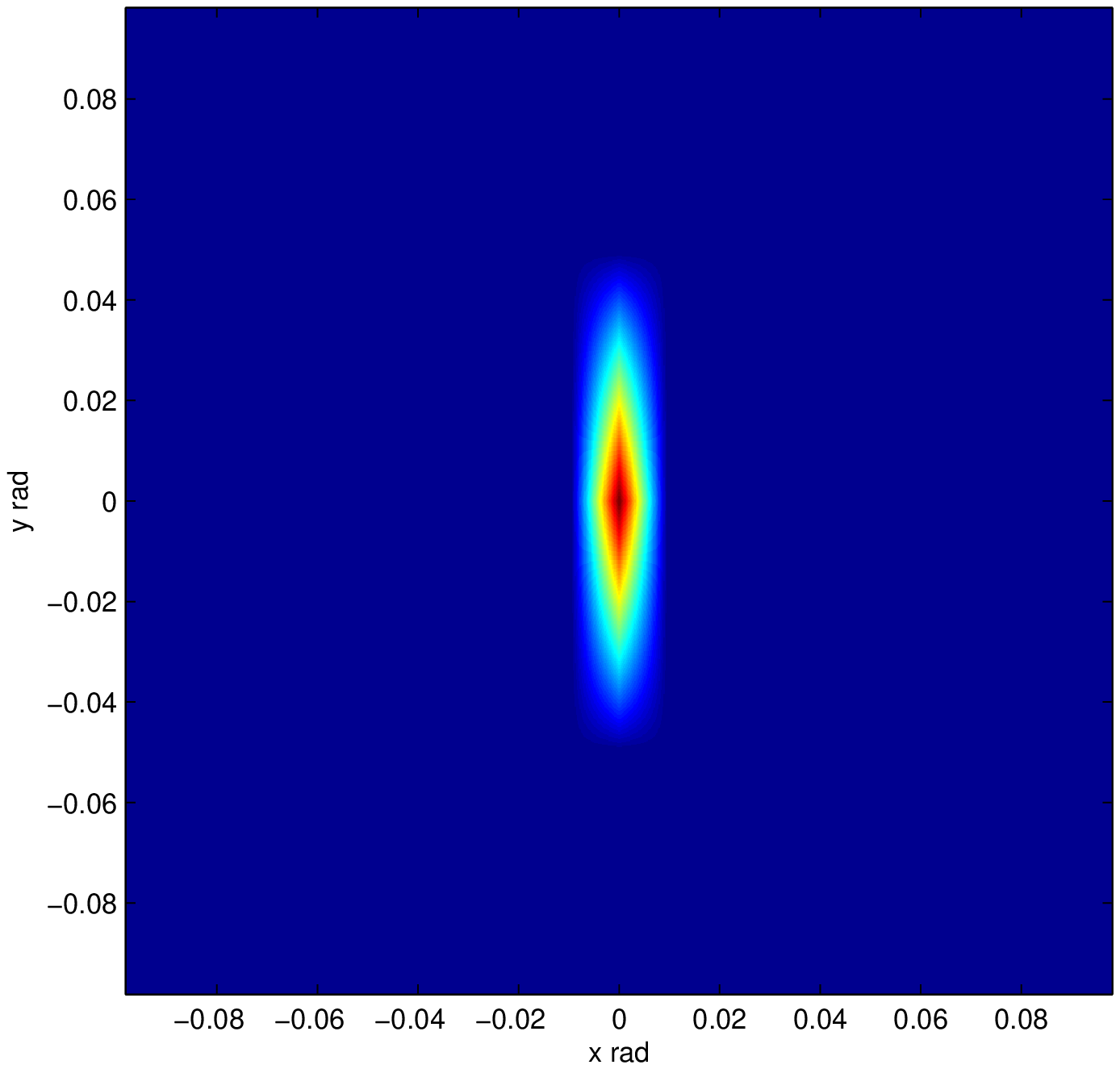}
\includegraphics[width=0.3\linewidth]{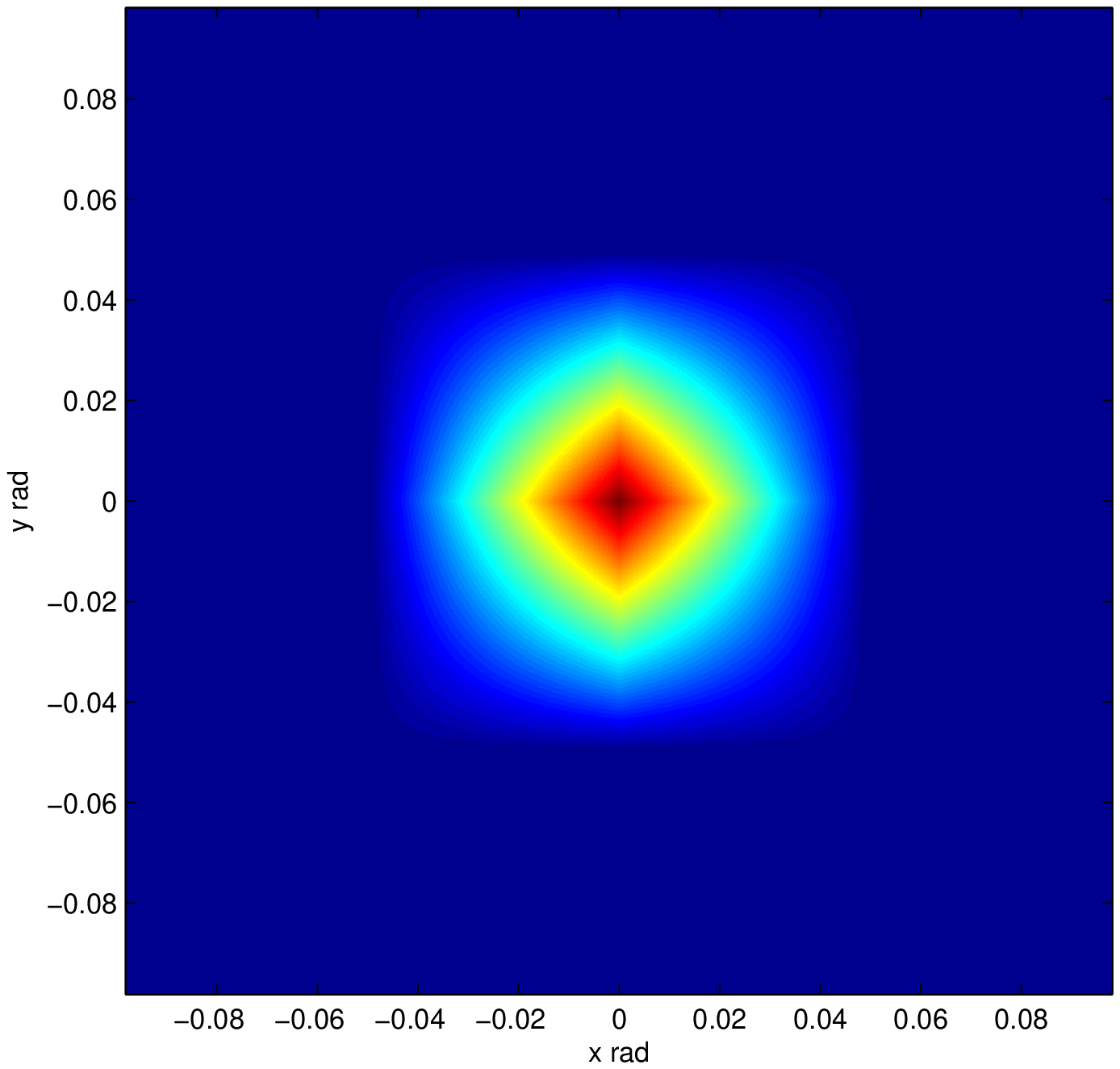}
\includegraphics[width=0.3\linewidth]{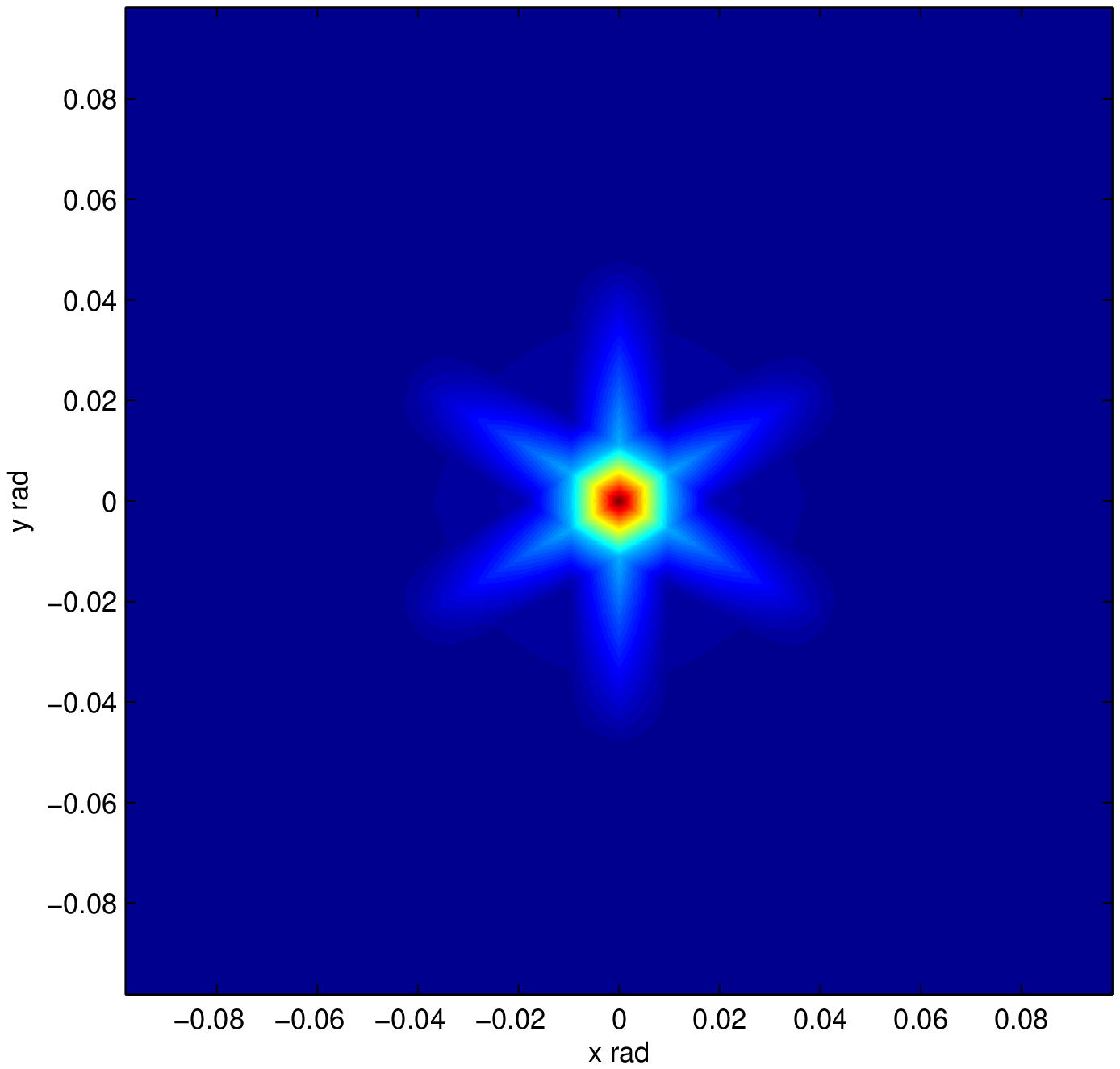}
\caption{PSFs of HXMT main detector. Left: PSF of collimator with $5.7^\circ \times 1^\circ$ FOV; middle: PSF of collimator with $5.7^\circ \times 5.7^\circ$ FOV; right: PSF of the high energe detector, overlaid.}
\label{fg-psf}
\end{figure}

The PSF of collimator $c$ is defined through its modulation kernel function as
\begin{equation}
P(\phi,\theta, c) = p(\phi,\theta,0,c,0,0)\text{,}
\end{equation}
i.e., the response of collimator $c$ to a unit point source located at the null position of the celestial sphere when the collimator is pointing at $(\phi,\theta)$ and its position angle parameter remains zero.
In practice the PSF is measured during the calibration of the detector\citep{han2010}, estimated from observations, or predicted theoretically.
The response to the unit object is determined only by the position of the unit object relative to the collimator rather than their absolute positions in any coordinate system.
So the modulation kernel function in a specific coordinate system can be evaluated through the PSF $P(\phi,\theta,c)$ by coordinate transforms.
For example, the PSF of collimators as well as the HE detector are shown in Fig. \ref{fg-psf}.

To evaluate the modulation kernel function $p(\phi,\theta,\psi,c,\phi',\theta')$ we find a coordinate system where the coordinate of the unit object is $\Bigr(\begin{smallmatrix}
1\\
0\\
0
\end{smallmatrix}\Bigl)$, i.e., the null position of the new coordinate system and the position angle parameter of the collimator yields zero.
Once we find the pointing parameters $\Phi$ and $\Theta$ of the collimator in the new coordinate system, the modulation kernel function can be evaluated immediately.

This is implemented by the following steps:
\begin{enumerate}
\item Rotate the current coordinate system $S_0$ to the first auxiliary coordinate system $S_1$ where the unit object lies on $\Bigr(\begin{smallmatrix}
1\\
0\\
0
\end{smallmatrix}\Bigl)$.
For example, we can first rotate the orginal coordinate $S_0$ around its $z$-axis by $\phi'$ then arount its $y$-axis by $-\theta'$.
\item Find the status parameters $\phi_1$, $\theta_1$, and $\psi_1$ in $S_1$.
The coordinate of a vector in the rotated coordinate system $S_1$ is equivalent to rotating the vector inversely, i.e., rotating it around the $z$-axis of $S_0$ by $-\phi'$ then rotating around $y$-axis of $S_0$ by $\theta'$.
We formulate the rotation by quaternions as:
\begin{align}
\boldsymbol{q}_1 & = \cos\frac{\phi'}{2} - \sin\frac{\phi'}{2}\mathrm{k}\text{,}\\
\boldsymbol{q}_2 & = \cos\frac{\theta'}{2} + \sin\frac{\theta'}{2}\mathrm{j}\text{,}\\
\boldsymbol{q} & = \boldsymbol{q}_2 \boldsymbol{q}_1\text{,}
\end{align}
where $\boldsymbol{q}_1$ and $\boldsymbol{q}_2$ are auxiliary quaternions.
The corresponding rotation matrix is
\begin{equation}
\boldsymbol{R}_{\theta',\phi'} = \begin{pmatrix}
  \cos\theta' \cos\phi' &   \cos\theta' \sin\phi' & \sin\theta' \\
-             \sin\phi' &               \cos\phi' &           0 \\
- \sin\theta' \cos\phi' & - \sin\theta' \sin\phi' & \cos\theta'
\end{pmatrix}\text{.}
\end{equation}
Given that the pointing vector and position vector of the collimator are $\boldsymbol{p}_{\phi,\theta,\psi}$ and $\boldsymbol{a}_{\phi,\theta,\psi}$ in $S_0$ as stated in Eq. \ref{eq-pointing} and Eq. \ref{eq-azimuthal} respecively,
the two vectors of the collimator are
\begin{equation}
\boldsymbol{p}_1 = \boldsymbol{R}_{\theta',\phi'}\boldsymbol{p}_{\phi,\theta,\psi}\text{,}
\end{equation}
and
\begin{equation}
\boldsymbol{a}_1 = \boldsymbol{R}_{\theta',\phi'}\boldsymbol{a}_{\phi,\theta,\psi}\text{.}
\end{equation}
\item Rotate $S_1$ around its $x$-axis by $-\alpha$ so that the position angle parameter of the given collimator in the rotated coordinate system $S_2$ is zero.
Given the pointing vector $\boldsymbol{p}_2$ and the position vector $\boldsymbol{a}_2$ of the collimator in $S_2$,
we have $(\boldsymbol{a}_2 \times \boldsymbol{p}_2) \cdot \mathrm{k} = 0$, i.e., the cross-product of the position vector and the pointing vector lies on the $x O y$ plane of $S_2$.
Therefore we have
\begin{equation}
\alpha = \operatorname{Arg}(\boldsymbol{R}_{\theta',\phi'}(\boldsymbol{a}_{\phi,\theta,\psi} \times \boldsymbol{p}_{\phi,\theta,\psi}) \cdot \mathrm{k},  \boldsymbol{R}_{\theta',\phi'}(\boldsymbol{a}_{\phi,\theta,\psi} \times \boldsymbol{p}_{\phi,\theta,\psi}) \cdot \mathrm{j})\text{,}
\label{eq-alpha}
\end{equation}
where $\operatorname{Arg}(x,y)$ is the principal argument of a complex number with $x + y\mathrm{i}$.
\item Find the pointing parameters $\phi_2$ and $\theta_2$ of the collimator in $S_2$ and use them to evaluate the PSF, which is equivalent to the modulation kernel function in $S_2$.
Given the rotation matrix
\begin{equation}
\boldsymbol{R}_\alpha = \begin{pmatrix}
1 &          0 &            0\\
0 & \cos\alpha & - \sin\alpha\\
0 & \sin\alpha &   \cos\alpha
\end{pmatrix}\text{,}
\end{equation}
where $\alpha$ is solved in Eq. \ref{eq-alpha}, the pointing vector $\boldsymbol{p}_2$ of the collimator is
\begin{equation}
\boldsymbol{p}_2 = \boldsymbol{R}_\alpha^{-1} \boldsymbol{R}_{\phi',\theta'} \boldsymbol{p}_{\phi,\theta,\psi}\text{.}
\end{equation}
Hence we can find the pointing parameters $\phi_2$ and $\theta_2$ from $\boldsymbol{p}_2$ as
\begin{align}
\phi_2 = & \operatorname{Arg}(\boldsymbol{p}_2 \cdot \mathrm{j}, \boldsymbol{p}_2 \cdot \mathrm{i})\text{,}\\
\theta_2 = & \arcsin(\boldsymbol{p}_2 \cdot \mathrm{k})\text{.}
\end{align}
Finally the modulation kernel function is evaluated as
\begin{equation}
p(\phi,\theta,\psi,c,\phi',\theta') = P(\phi_2,\theta_2,c)\text{.}
\end{equation}
\end{enumerate}

\subsubsection{Numerical evaluation of modulation}
Since the response of a detector of HXMT to an object depends upon the position of the object relative to the detector only,
we can always rotate both the detector and the object to the initial tessella of the celestial sphere to calculate the modulation kernel function,
if the tessella is larger than the FOV of the detector, i.e.,
\begin{equation}
p(\boldsymbol{p}_{\phi,\theta,\psi},\boldsymbol{a}_{\phi,\theta,\psi},c,\boldsymbol{p}_{\phi',\theta'}) = 
p(\boldsymbol{R}_{\phi_w, \theta_w}^{-1}\boldsymbol{p}_{\phi,\theta,\psi},
\boldsymbol{R}_{\phi_w, \theta_w}^{-1}\boldsymbol{a}_{\phi,\theta,\psi},c,
\boldsymbol{R}_{\phi_w, \theta_w}^{-1}\boldsymbol{p}_{\phi',\theta'})\text{,}
\end{equation}
where $\boldsymbol{p}_{\phi,\theta,\psi}$ and $\boldsymbol{a}_{\phi,\theta,\psi}$ represent the pointing vector and the position vector of the collimator which are related to its status $(\phi,\theta,\psi)$, $\boldsymbol{p}_{\phi',\theta'}$ represents the point $(\phi',\theta')$, so $p(\boldsymbol{p}_{\phi,\theta,\psi},\boldsymbol{a}_{\phi,\theta,\psi},c,\boldsymbol{p}_{\phi',\theta'})$ is equivalent to $p(\phi,\theta,\psi,c,\phi',\theta')$.
Rotation matrix $\boldsymbol{R}_{\phi_w,\theta_w}$ rotates the initial tessella to $(\phi_w, \theta_w)$.

We use the first-order terms of its Taylor series to approximate the modulation kernel function around $\phi=0$, $\theta=0$, $\phi'=0$ and $\theta'=0$ (i.e., on the initial tessella), as
\begin{equation}
p(\phi,\theta,\psi,c,\phi',\theta') = (\boldsymbol{R}_{\psi} P)(\phi - \phi', \theta - \theta', c) + O(\phi^2) + O(\theta^2) + O(\phi'^2) + O(\theta'^2)\text{,}
\end{equation}
where we put a rotation matrix $\boldsymbol{R}_{\psi}$ to the left of the PSF $P$ to define a new PSF $(\boldsymbol{R}_{\psi} P)$ which is rotated from the original one.
$O(\phi^2)$, $O(\theta^2)$, $O(\phi'^2)$ and $O(\theta'^2)$ are the remainders.

Taking the pixelization of observed data into account, the corresponding discrete modulation equation yields
\begin{equation}
d_{i,j,c} = \sum_{i',j'}^N (\boldsymbol{R}_{\psi_{i,j}} P)_{i-i', j-j', c} f_{i',j'}\text{,}
\label{eq-modulation-approx}
\end{equation}
where $\psi_{i,j}$ is the position angle parameter of the collimator $c$ on $(i,j)$ and $P_{i,j,c}$ is the value of the PSF of the collimator on $(i,j)$.

\subsubsection{Position angle cluster analysis and approximation}
\begin{figure}[htbp]
\centering
\includegraphics[height=35ex]{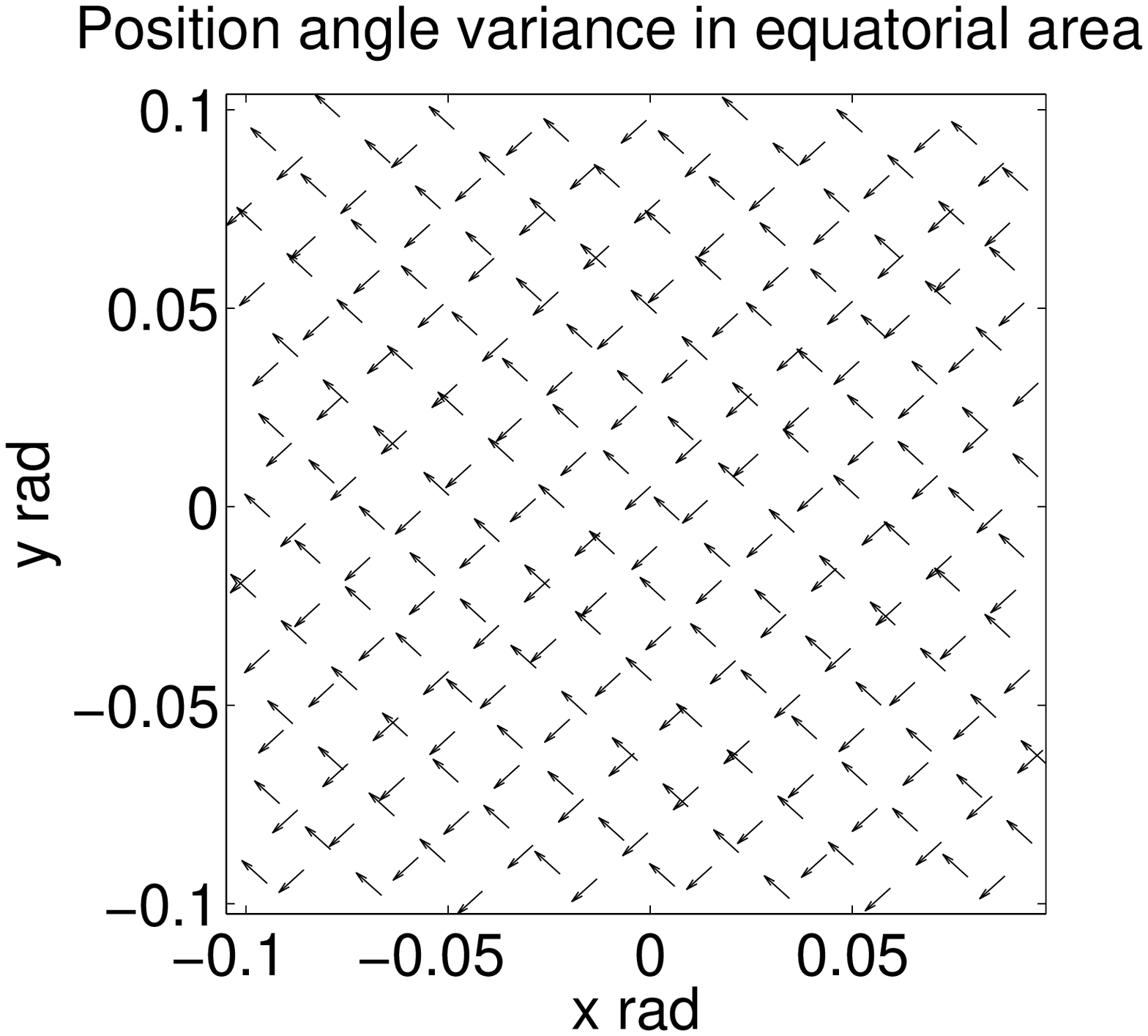}
\includegraphics[height=35ex]{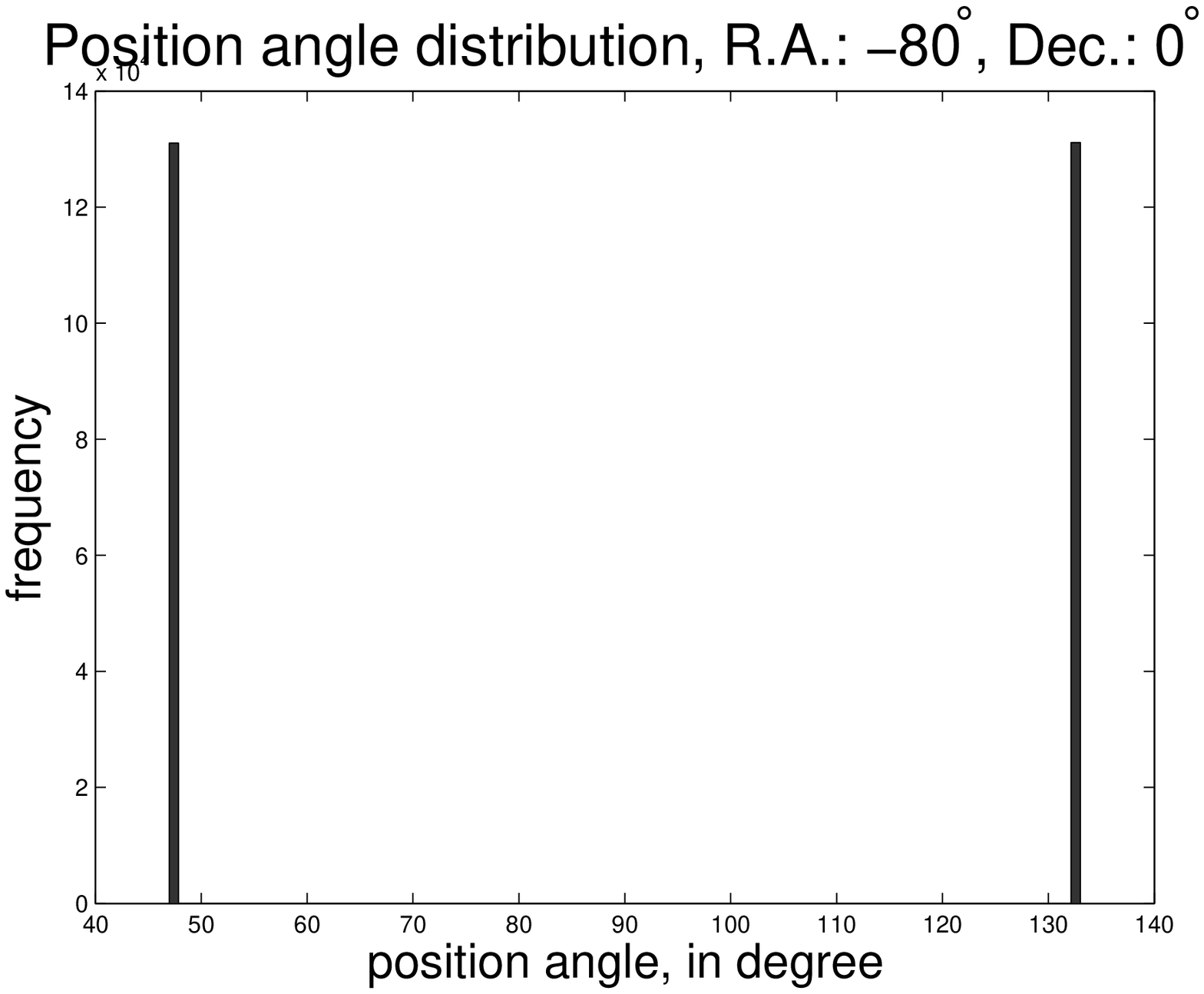}\\
\includegraphics[height=35ex]{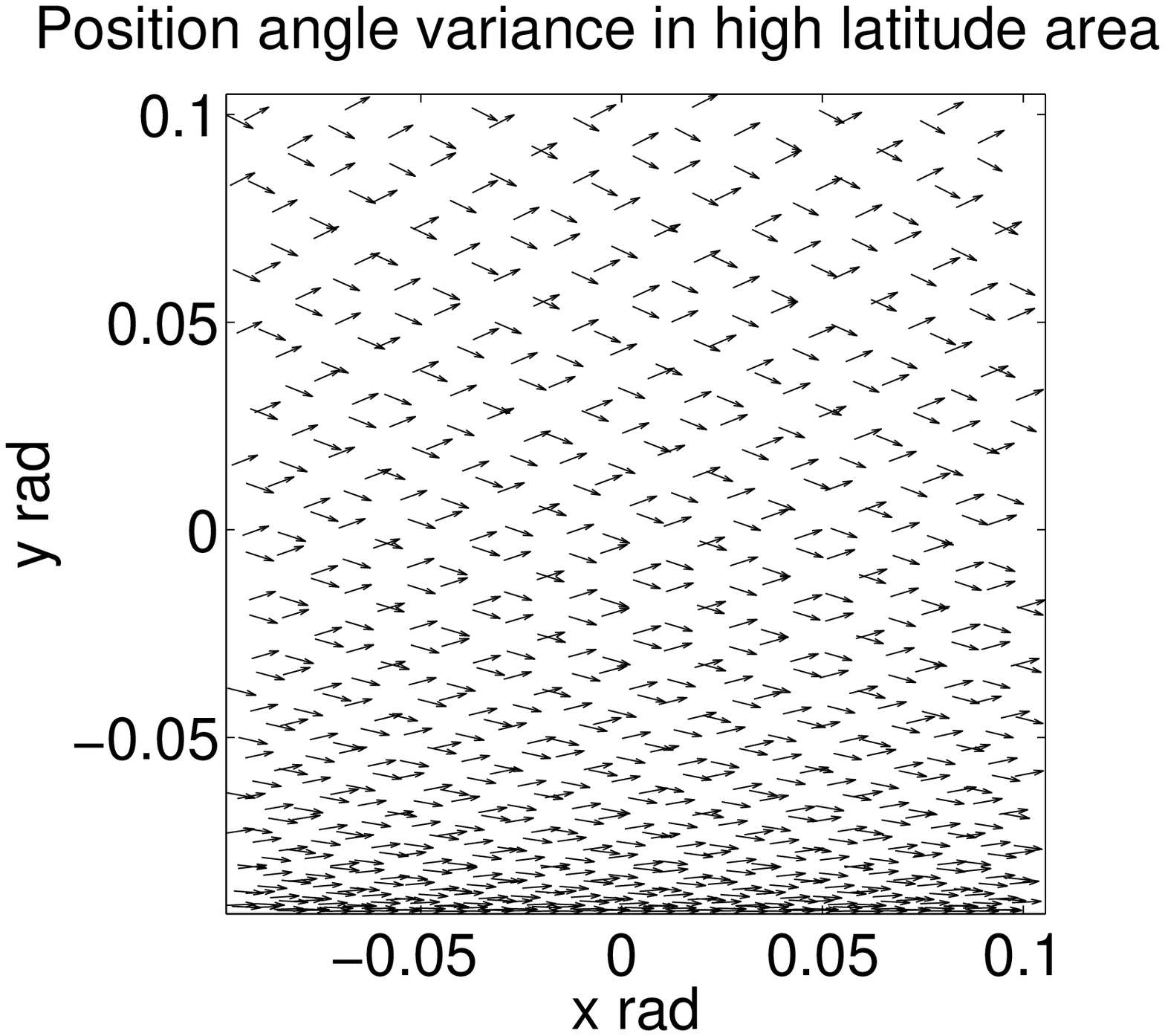}
\includegraphics[height=35ex]{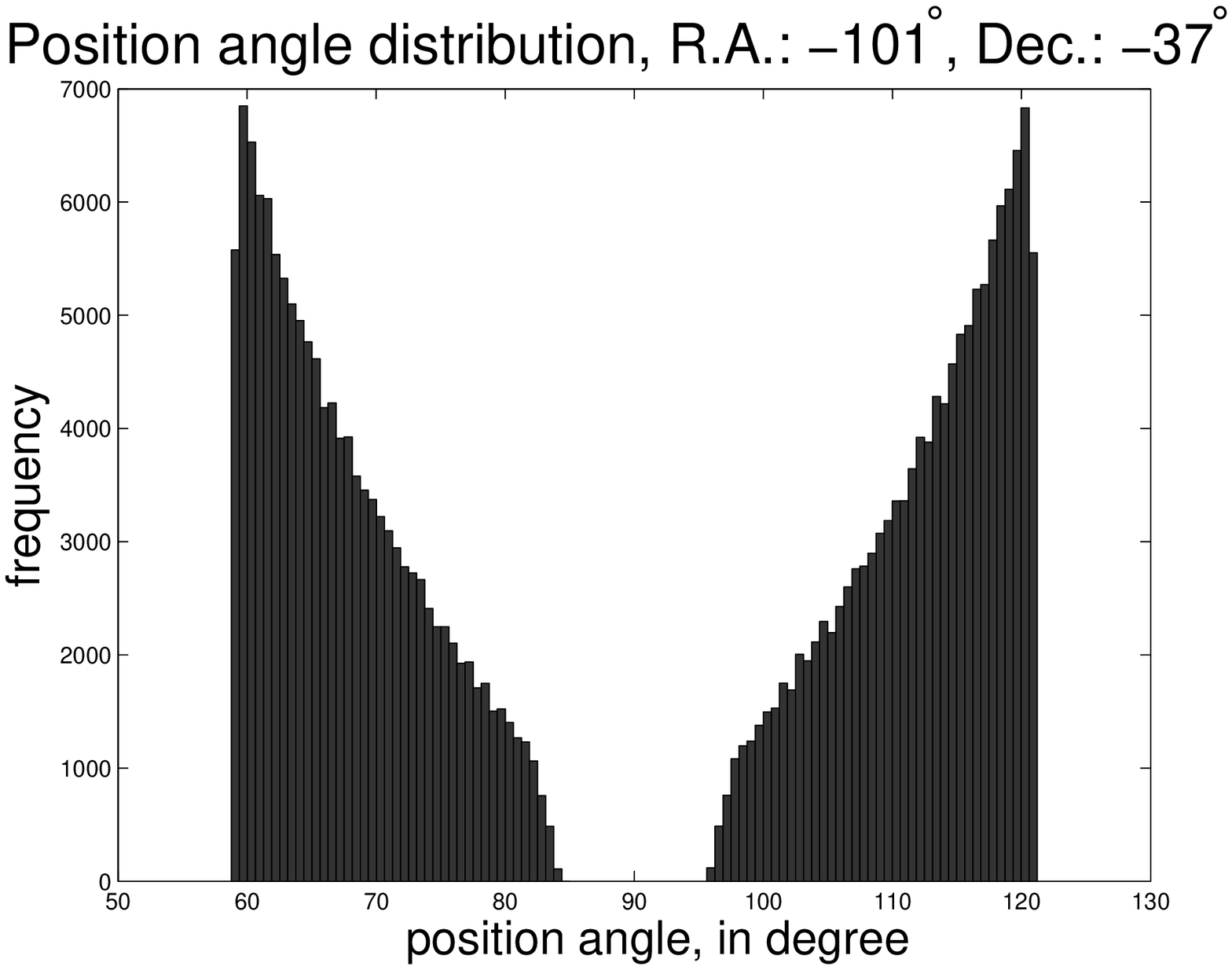}
\caption{Position angle variance and distribution. \textbf{Top left:} position angle variance along scanning path of all-sky survey in equatorial area; \textbf{top right:} position angle distribution in equatorial area; \textbf{bottom left:} positon angle variance along scanning path of all-sky survey in high latitude area; \textbf{bottom right:} position angle distribution in high latitude area.}
\label{fg-azimuth-variance}
\end{figure}

Because of the topology of the spherical surface, the position angle parameter of each collimator varies during the all-sky survey and the variance depends upon the position on the celestial sphere.
Given a specific scanning scheme for the all-sky survey, e.g., the satellite orbits along the path shown in Fig. \ref{fg-scanning-path} and its roll angle is fixed on a constant value such as $0^\circ$, $-30^\circ$ or $30^\circ$,
the position angle parameter of a collimator is either $\psi_c - 43^\circ$ or $\psi_c + 43^\circ$ ($\psi_c$ is the angle between the position vector of the collimator and the instantaneous velocity of the satellite, i.e., the tangent vector of the scanning trajectory, and the inclination of the satellite orbit is $43^\circ$\cite{li2007}) approximately in the initial tessella,
but varies in a wide range significantly in some other tesselae, as shown in Fig. \ref{fg-azimuth-variance}.

On the pixel grid of a small sky region we can rotate the PSF around itself and shift it along two orthogonal directions to evaluate the modulation kernel function approximately, as shown in Equation \ref{eq-modulation-approx}.
Because of the position angles variance as well as the lack of circular symmetry of the PSF the modulation kernel function is shift-variant.
We can approximate the modulation kernel function with the sum of a finite series of functions so that the position angle parameter of the collimator status is fixed for each function, as
\begin{equation}
\begin{split}
p(\phi,\theta,\psi,c,\phi',\theta) 
& =       \int_{-\pi}^{\pi} p(\phi,\theta,\psi',c,\phi',\theta')\delta(\psi - \psi')\mathrm{d}\psi'\\
& \approx \sum_{k=1}^{K}    p\biggl(\phi,\theta,\frac{2 \pi k}{K}-\pi,c,\phi',\theta'\biggr)\delta_{k,\frac{\psi+\pi}{2\pi}K}
\end{split}\text{,}
\end{equation}
where $\frac{\psi+\pi}{2\pi}K$ is rounded to the nearest integer to evaluate the Kronecker delta $\delta_{k,\frac{\psi+\pi}{2\pi}K}$ equivalent to the Dirac delta function $\delta(\psi - \psi')$.

We use cluster analysis to assign the position angle parameters of all the observed data in a given sky region into groups so that the difference between each position angle parameter of a group and the clustering center of the group is less than a pre-determined upper limit $\epsilon$ according to the required precision.
Therefore the modulation kernel function is shift-invariant approximately if all of its possible position angle parameters are in the same group.
So we rewrite Equation \ref{eq-modulation-approx} by expanding the right-hand side:
\begin{equation}
\begin{split}
d_{i,j,c} 
& = \sum_{i',j'}^N (\boldsymbol{R}_{\psi_{i,j}} P)_{i-i', j-j', c} f_{i',j'}\\
& = \sum_{k=1}^{K} \sum_{i',j'}^N (\boldsymbol{R}_{\psi_k} P)_{i-i', j-j', c} f_{i',j'} \delta(|\psi_k - \psi_{i,j}|<\epsilon)
\end{split}\text{,}
\end{equation}
where
\begin{equation}
\delta(|\psi_k - \psi_{i,j}|<\epsilon) = \begin{cases}
& 1\quad\text{if}\;|\psi_k - \psi_{i,j}|<\epsilon\\
& 0\quad\text{otherwise}
\end{cases}\text{.}
\end{equation}
Finally we approximate the modulation by the sum of a finite series of convolution as
\begin{equation}
d_{i,j,c} = \sum_{k=1}^{K} [(\boldsymbol{R}_{\psi_k} P_c) \ast f]_{i,j} \delta(|\psi_k - \psi_{i,j}|<\epsilon)\text{,}
\end{equation}
where $[(\boldsymbol{R}_{\psi_k} P_c) \ast f]$ represents the convolution between the rotated PSF of collimator $c$ and the image $f$.
Then we can employ FFT algorithms to accelerate the modulation.

\section{Simulated data and reconstructed images}
\begin{figure}[htbp]
\centering
\includegraphics[width=0.9\linewidth]{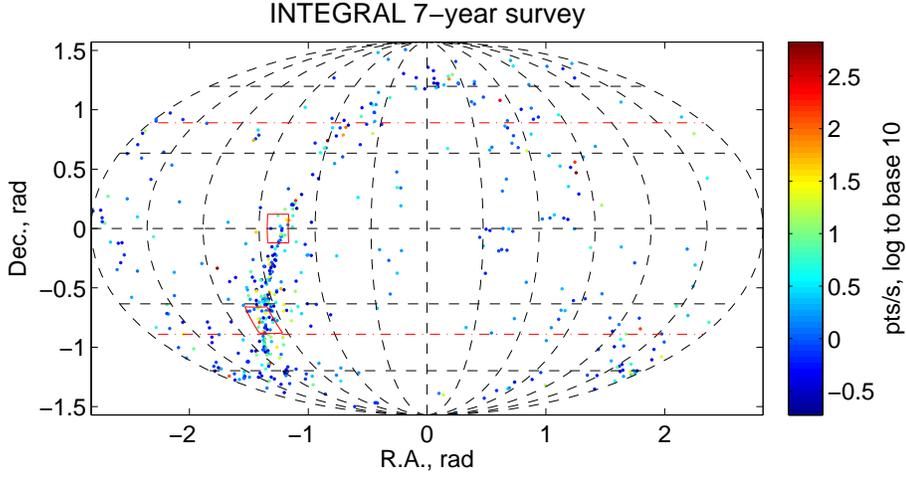}
\caption{Sources in 7-year INTEGRAL all-sky survey catalog.
\textbf{Upper box:} a tessella in the equatorial area.
The R.A. and Dec. of its center are $-80^\circ$ and $0^\circ$ respectively.
\textbf{Lower box:} a tessella in the high latitude area.
The R.A. and Dec. of its center are $-101^\circ$ and $-37^\circ$.}
\label{fg-obj}
\end{figure}

\begin{figure}[htbp]
\centering
\includegraphics[width=0.48\linewidth]{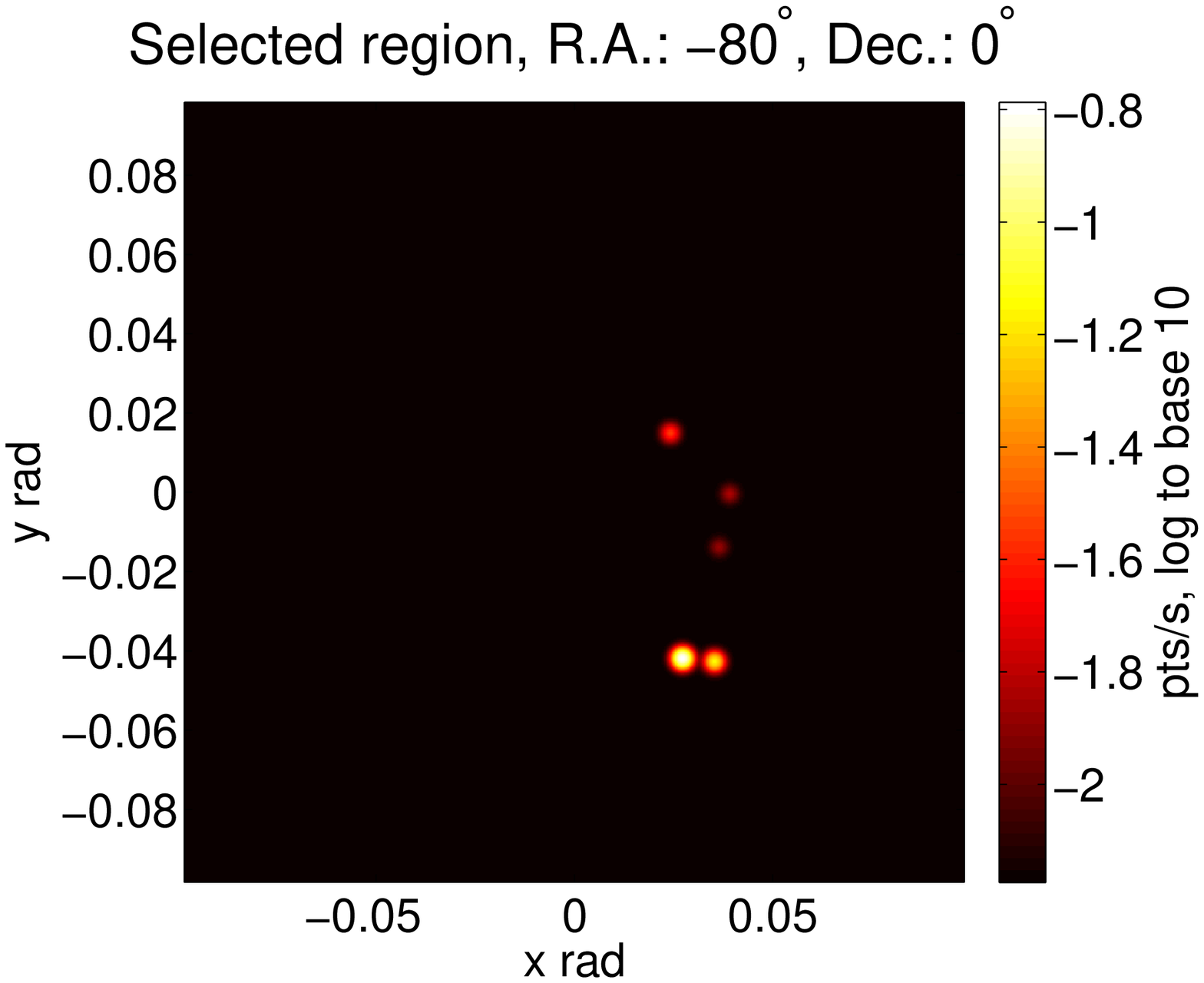}
\includegraphics[width=0.48\linewidth]{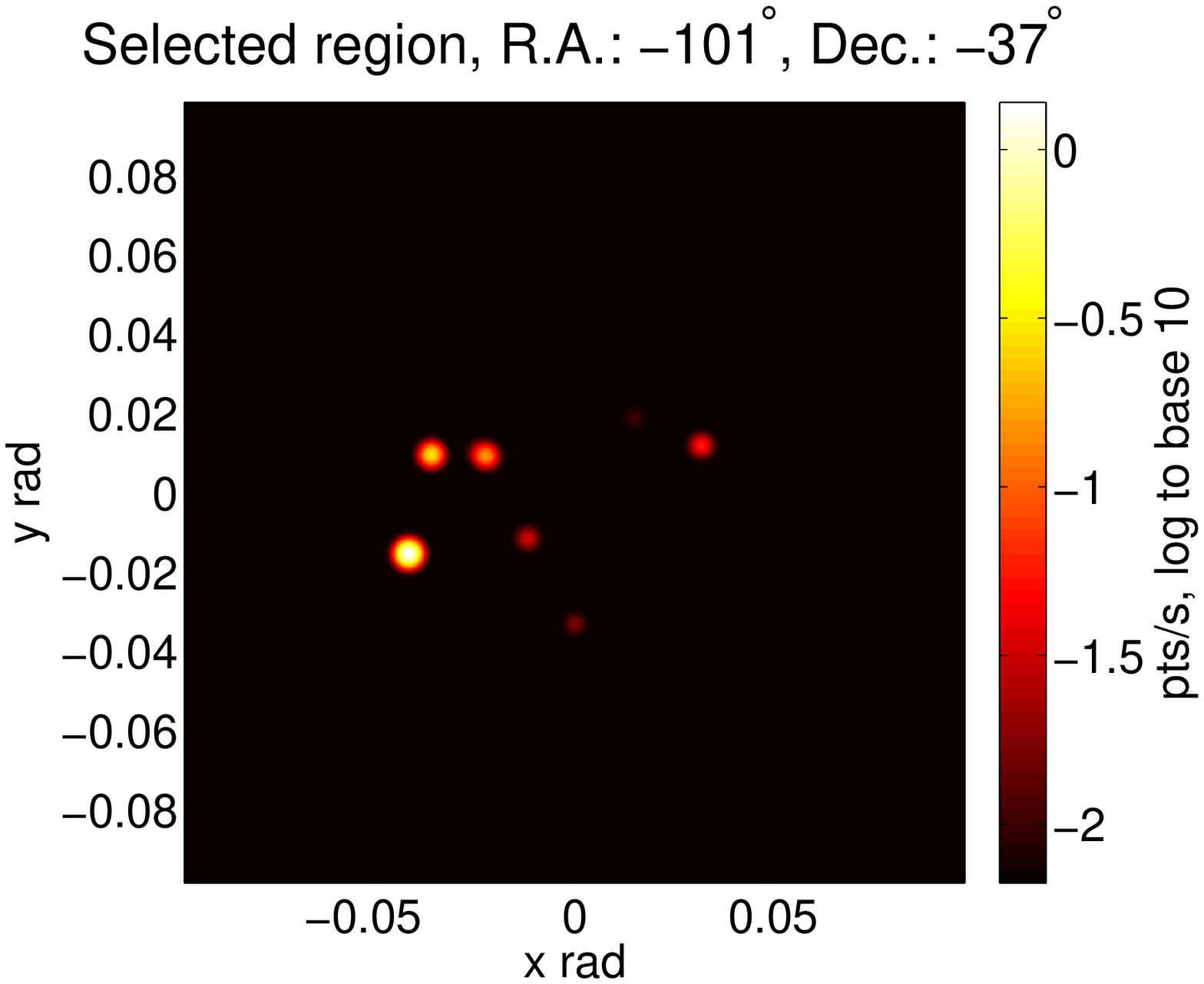}\\
\includegraphics[width=0.48\linewidth]{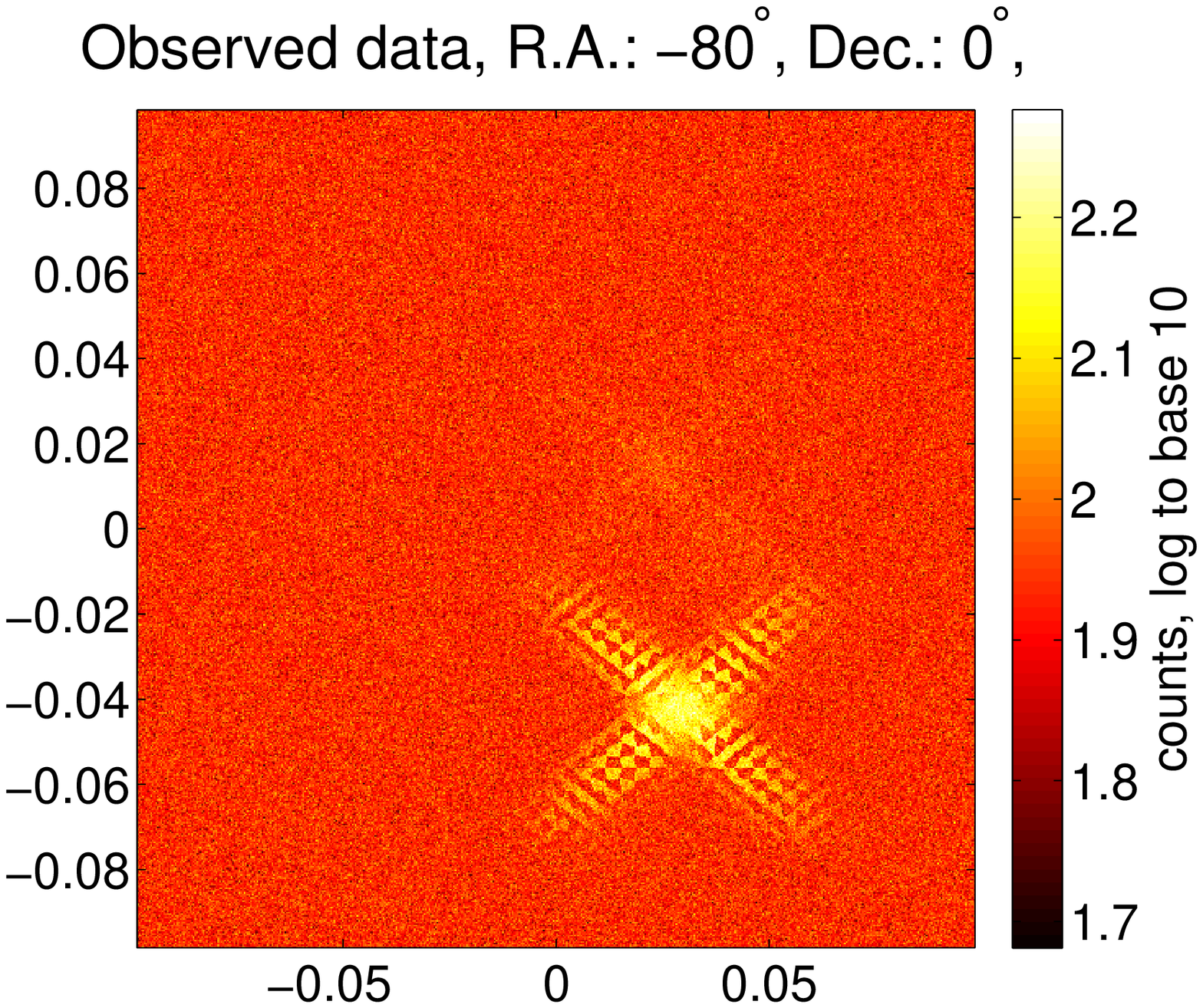}
\includegraphics[width=0.48\linewidth]{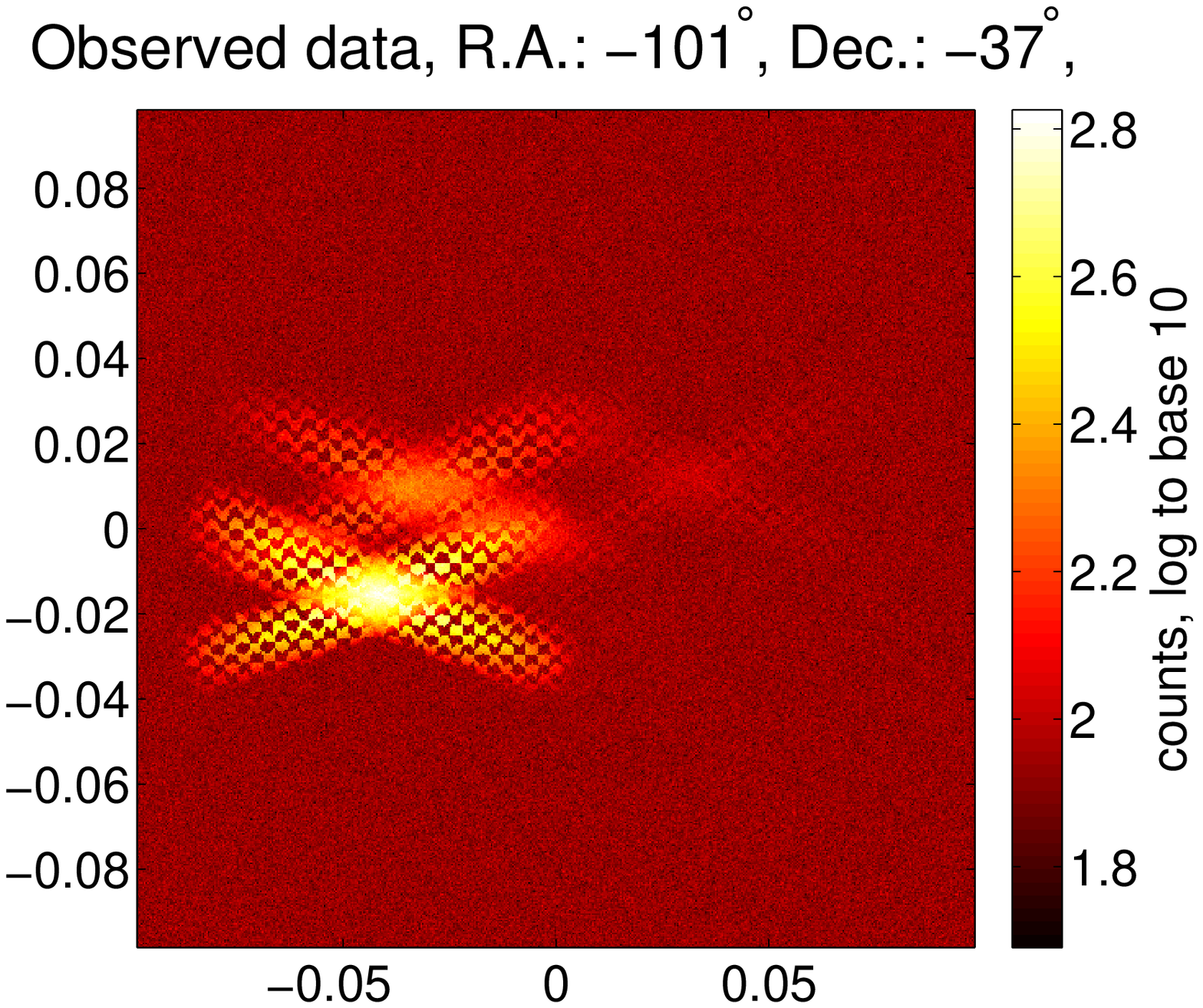}\\
\includegraphics[width=0.48\linewidth]{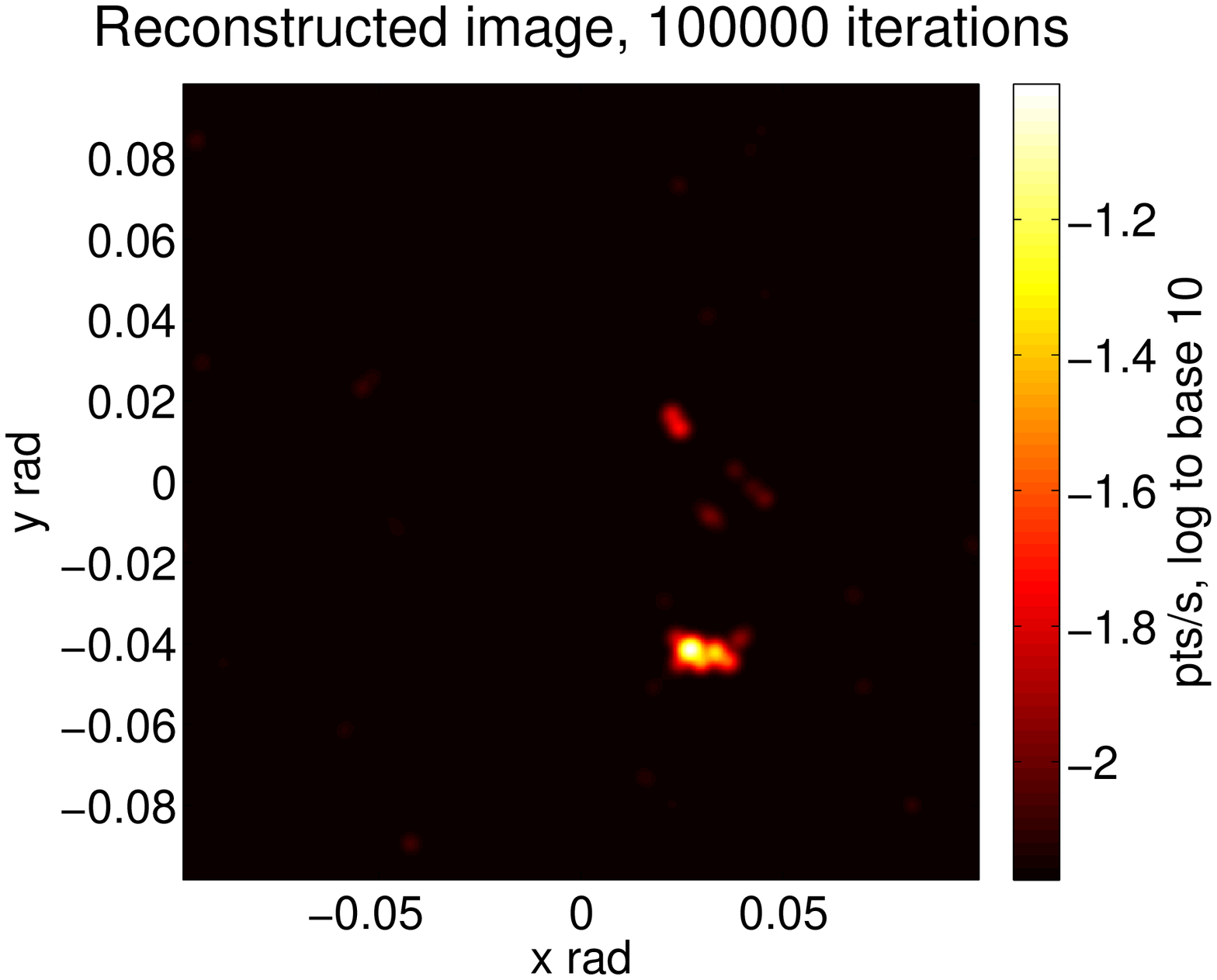}
\includegraphics[width=0.48\linewidth]{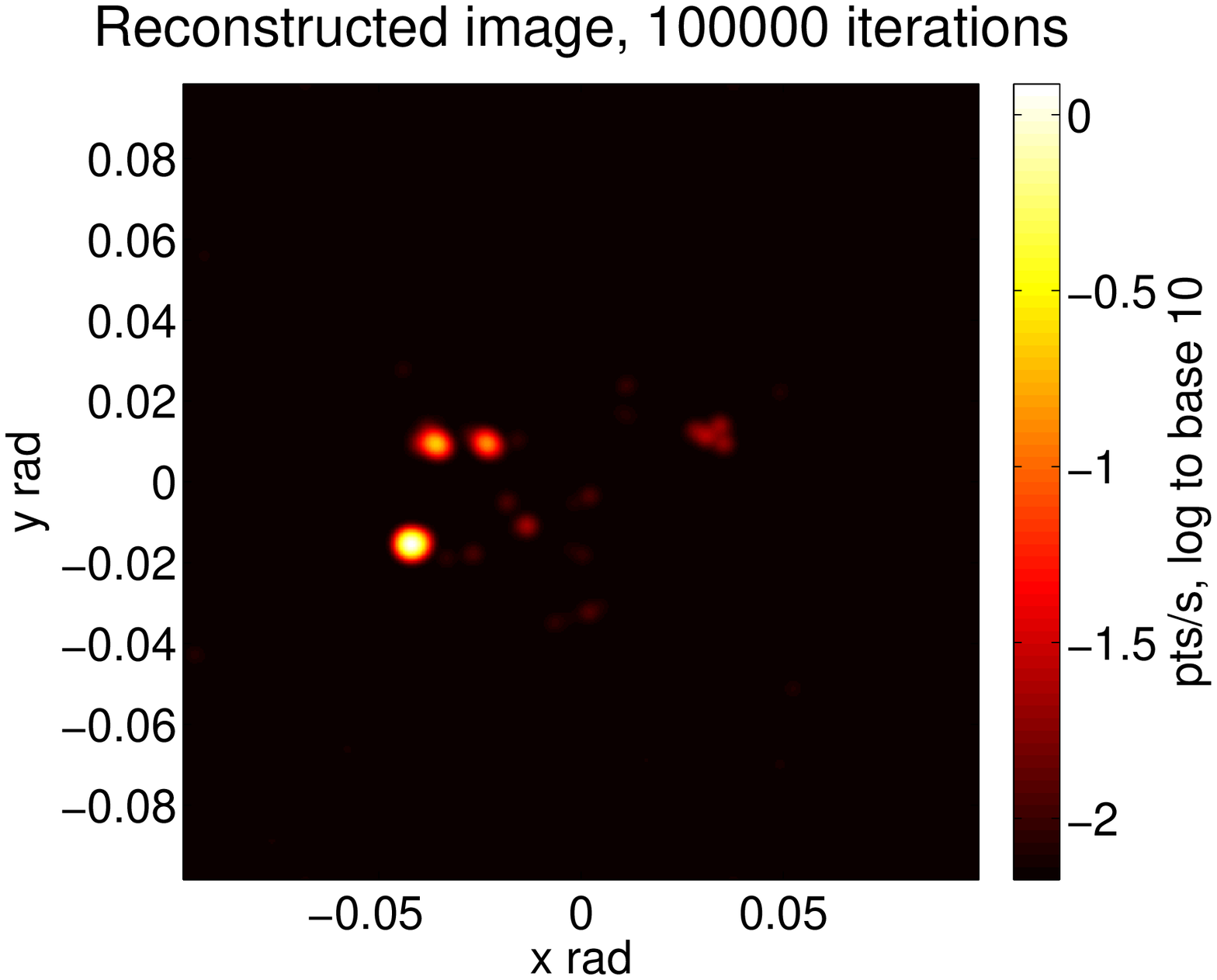}
\caption{\textbf{Top left:} Model image in equatorial area.
\textbf{Top right:} Model image in high latitude area.
\textbf{Middle left:} Observed image in equatorial area, with poisson noise.
\textbf{Middle right:} Observed image in high latitude area, with poisson noise.
\textbf{Bottom left:} Reconstructed image using accelerated DD method, with $100\;000$ iterations (about $3$ minutes).
\textbf{Bottom right:} Reconstructed image using accelerated DD method, with $100\;000$ iterations (about $10$ minutes).
}
\label{fg-obs-ddm}
\end{figure}

The 7-year INTEGRAL all-sky survey catalog \citep{integral7yr} is used as input to simulate HXMT data, as shown in Fig. \ref{fg-obj}.
Two small regions are selected for simulated observations as well as reconstructions.
One of them is in equatorial area (the model image is shown on top left of Fig. \ref{fg-obs-ddm}) while the other is in high latitude area (as shown on top right of Fig. \ref{fg-obs-ddm}).
A background of $0.01\;\mathrm{mCrab}$ at $20\;\mathrm{keV}$ is simulated.
The PSF of HE is shown in Fig. \ref{fg-psf}.
The total effective area of HE detectors corresponding to each position is $1\,500\;\mathrm{cm}^2$.
The exposure towards a $11.25^\circ \times 11.25^\circ$ region in equatorial area lasts $1.6 \times 10^4 \mathrm{s}$ and $3.0 \times 10^4 \mathrm{s}$ in high latitude area.
Hence the background level of observed data is $100\;\mathrm{cts}$ and $200\;\mathrm{cts}$ in equatorial area and in high latitude area respectively.
Poisson noise is added thus the standard deviation of the noise is about $10\;\mathrm{cts}$ and $14\;\mathrm{cts}$ in equatorial area and in high latitude area.
The position angle variance in high latitude area is more significant thus the simulated observation as well as the reconstruction consume more computation time.

The model images, observed data as well as the reconstructed images are shown in Fig. \ref{fg-obs-ddm}.
Fig. \ref{fg-obs-ddm} shows that with the accelerated DD we can obtain expected images from observed data,
thus the proposed method is valid.

To evaluated the performance of the method, the time and memory consumptions for different setups are shown in Table \ref{tb-compcost}.
\begin{table}[htbp]
\centering
\begin{tabular}{p{6em}|p{4em}|p{4em}|r|p{5em}|p{6em}|p{6em}}
                           & Time, in seconds                 & Memory, in bytes                  & Iterations & Size, in pixels         & Hardware performance, in GFlops & Algorithm 
acceleration\\ \hline
Equatorial area on CPU     & $3$ &    $1 \times 10^7$      & $100$ & $512 \times 512$  & $96$   & $30\,776$ \\ \hline
High-latitude area on CPU  & $9$ &    $2 \times 10^7$      & $100$ & $512 \times 512$  & $96$   & $10\,259$    \\ \hline
Equatorial area on GPU     & $0.2$    & $1 \times 10^7$    & $100$ & $512 \times 512$  & $1000$    & $44\,317$   \\ \hline
High-latitude area on GPU  & $0.6$    & $2 \times 10^7$    & $100$ & $512 \times 512$  & $1000$    & $14\,772$    \\ \hline
Original DD                & $1\,440$ & $2.8 \times 10^9$ & $100$ & $121 \times 121$  & $19.2$ & -
\end{tabular}
\caption{Computational costs of accelerated DD method and the original DD iterations.
The algorithm acceleration is independent of hardware performance or the size of the problem.
It is calculated as $(T_0 / T_1) \cdot (P_0 / P_1)$, where $T_0$ and $T_1$ are time costs of solving a problem in the same size by original DD and by the accelerated method respectively, $P_0$ and $P_1$ are the performances of the computers where original DD and the accelerated method are implemented.
Since the complexity of original DD is $O(n^2)$ ($n$ is the number of pixels),
the time cost of $100$ iterations on $512 \times 512$ pixels is about $1\,440\;\mathrm{s} \times \left(512 / 121\right)^{4} = 461\,637\;\mathrm{s}$.
For example, the algorithm acceleration for equatorial area on CPU is $(461\,637 / 3) \times (19.2/96) = 30\,776$.}
\label{tb-compcost}
\end{table}
The time costs of accelerated DD on CPU is measured on an Intel Core i7-2720QM with single process. But considering its \emph{AVX} feature (\emph{advanced vector extension}) as well as the fast fourier transform is implemented using FFTW library with \emph{pthread} (\emph{POSIX Threads}) enabled, multiple CPU cores are involved in a single process.
Its theoretical multi-core performance is $96\;\mathrm{GFlops}$.
The time costs on GPU is measured on an Nvidia Quadro 1000M GPU and projected to a more realistic desktop GPU with about $1\;\mathrm{TFLOPs/s}$ processing power.
The computational costs of orginal DD iterations are also included in Table \ref{tb-compcost} for comparison\citep{shen2007}.
The original DD was tested in 2007 on an Intel Core 2 Duo E6600 CPU.
Its theoretical multi-core performance is $19.2\;\mathrm{GFlops}$.
The accelerated and original DD have been tested on different systems and the mainstream computing power has been improved,
therefore we take not only the time costs but also the hardware performances into account.
The converted algorithm acceleration is also shown in Table \ref{tb-compcost}, which is hardware independent.

\begin{figure}[htbp]
\centering
\includegraphics[width=0.48\linewidth]{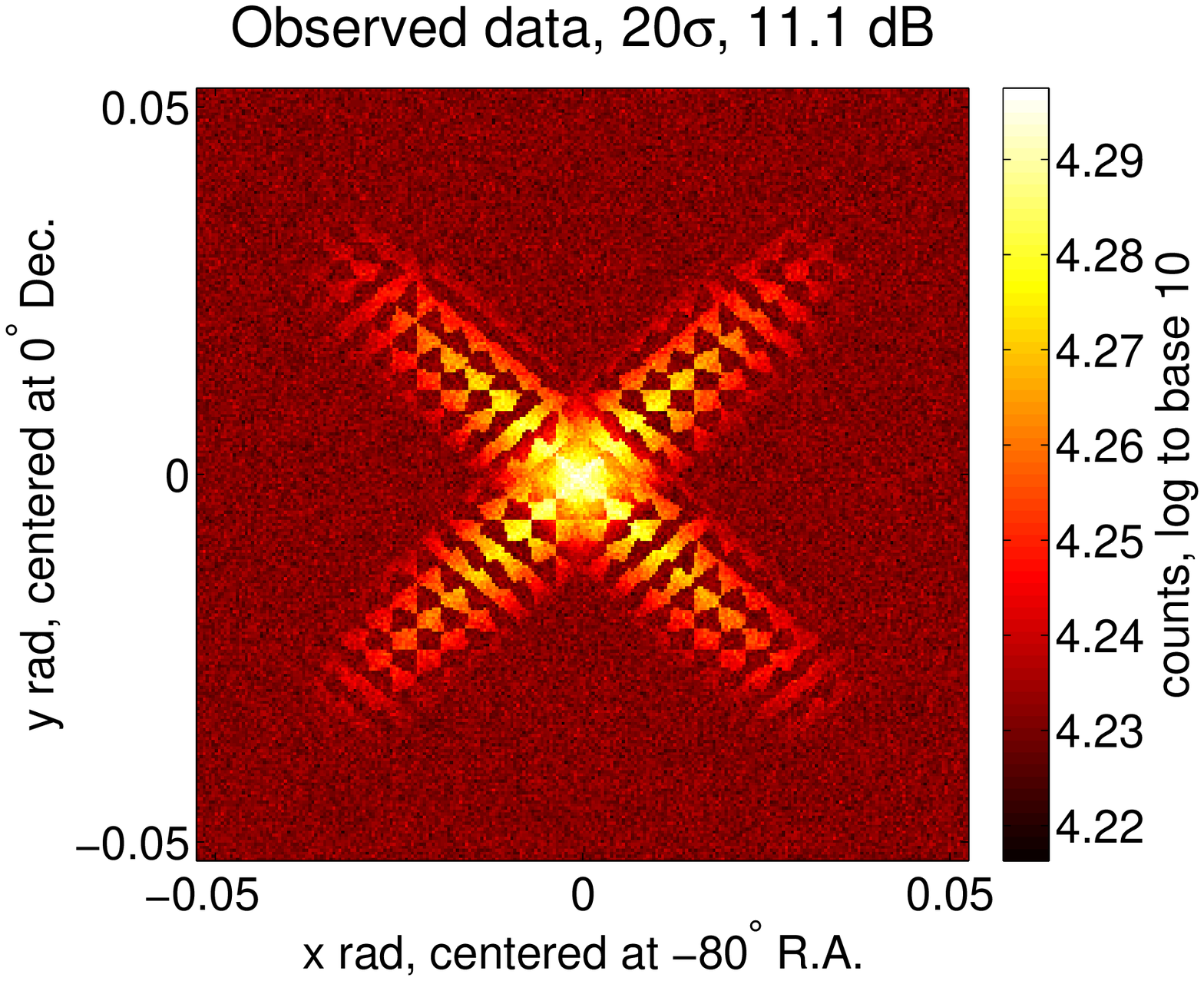}
\includegraphics[width=0.48\linewidth]{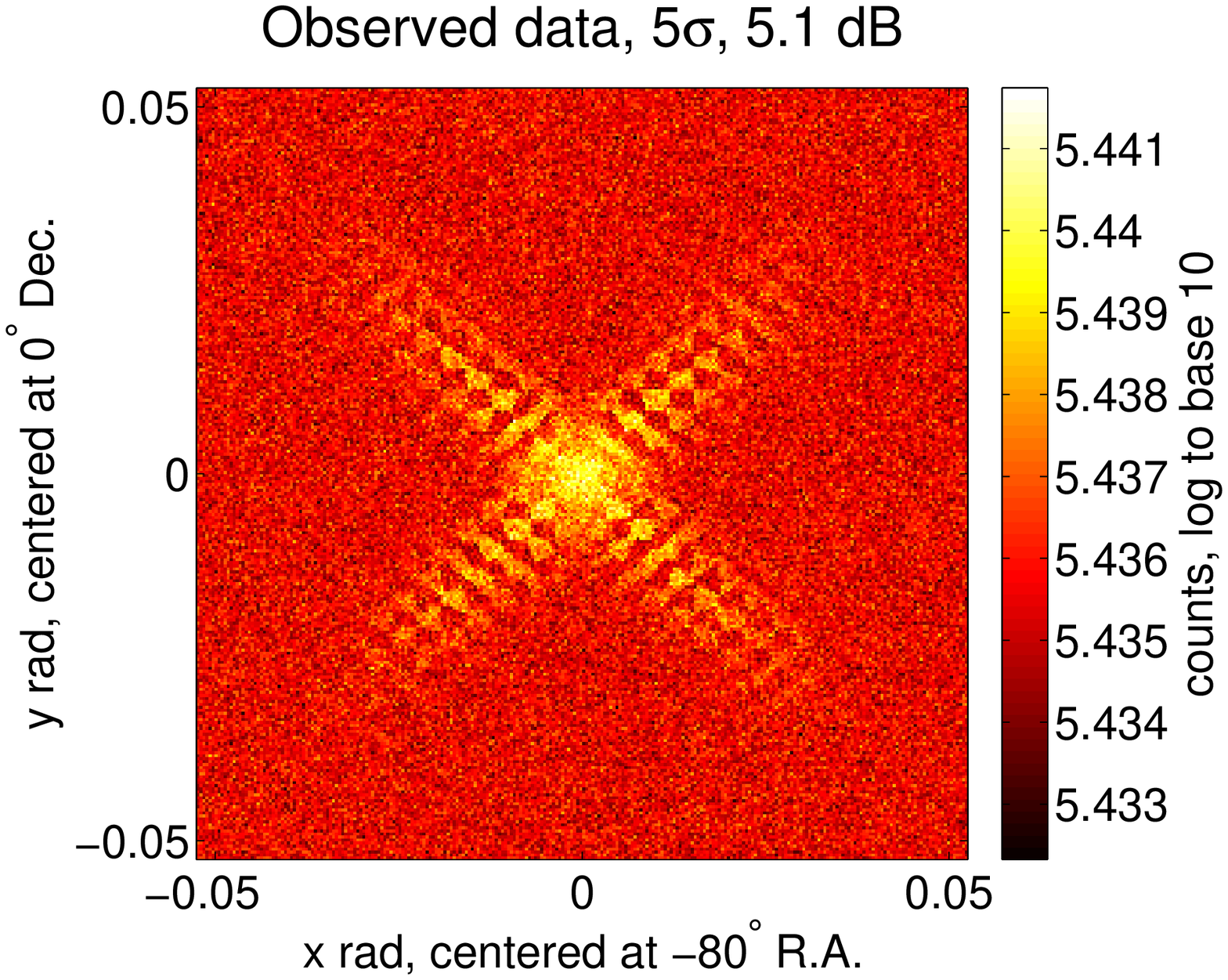}\\
\includegraphics[width=0.48\linewidth]{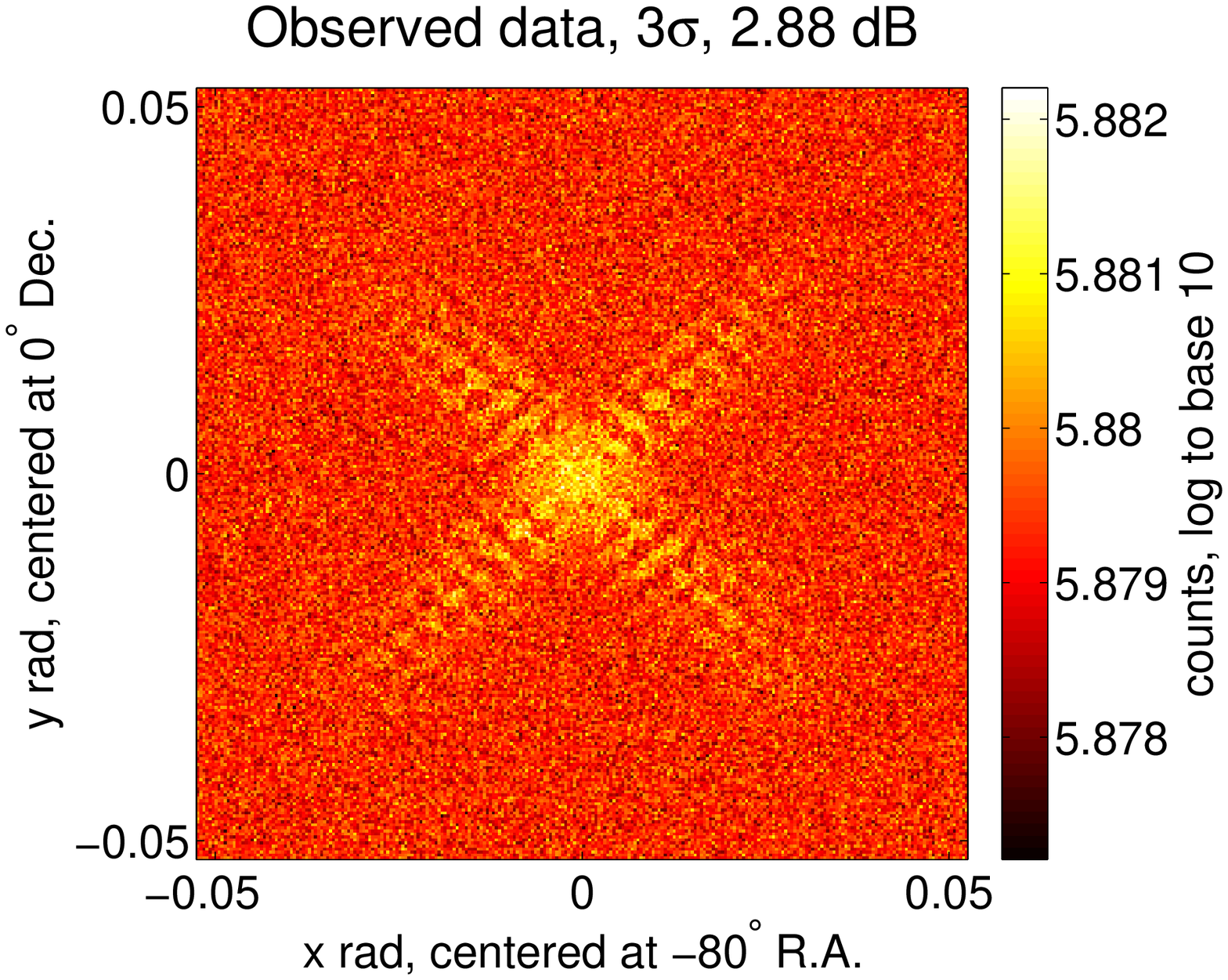}
\includegraphics[width=0.48\linewidth]{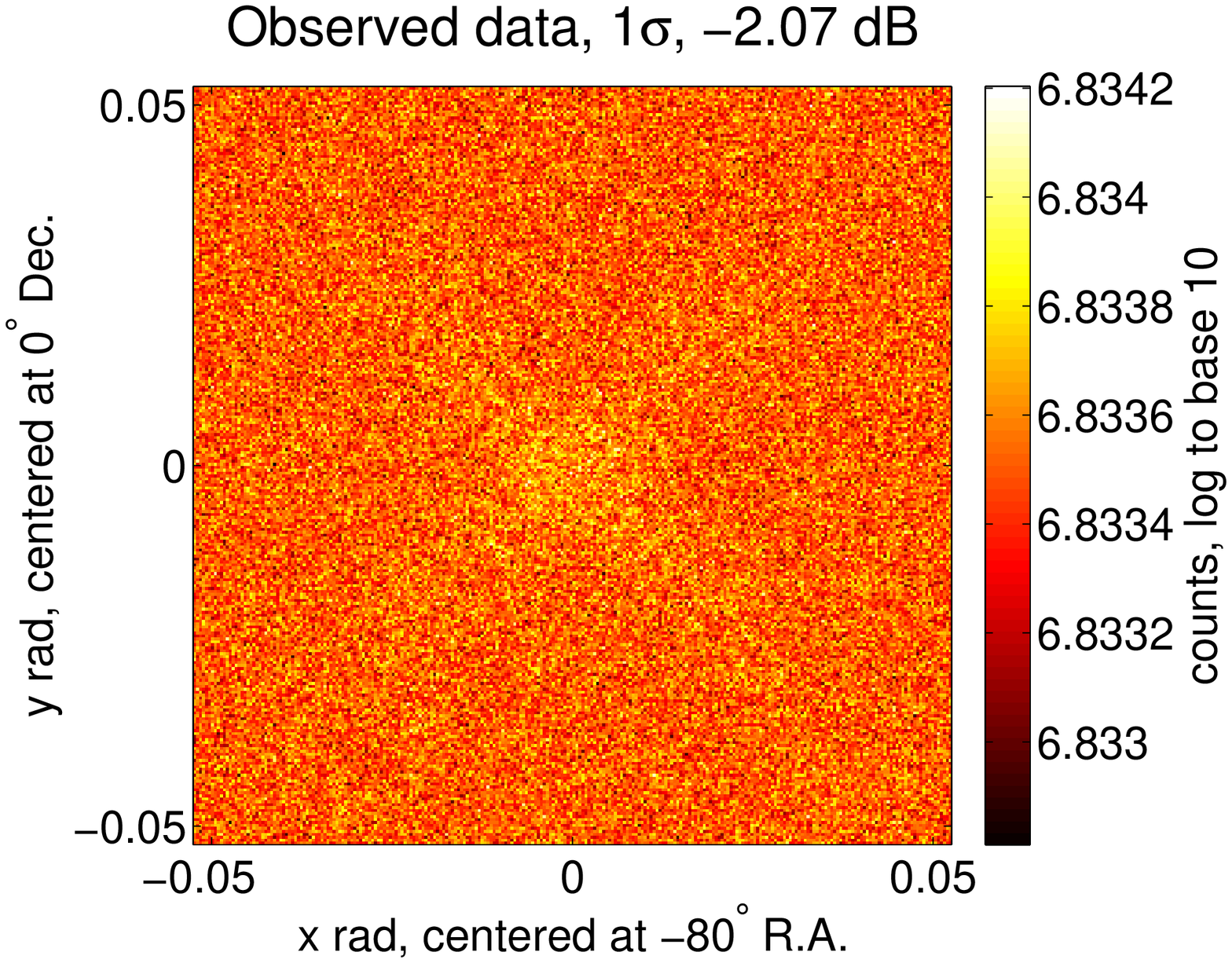}\\
\includegraphics[width=0.48\linewidth]{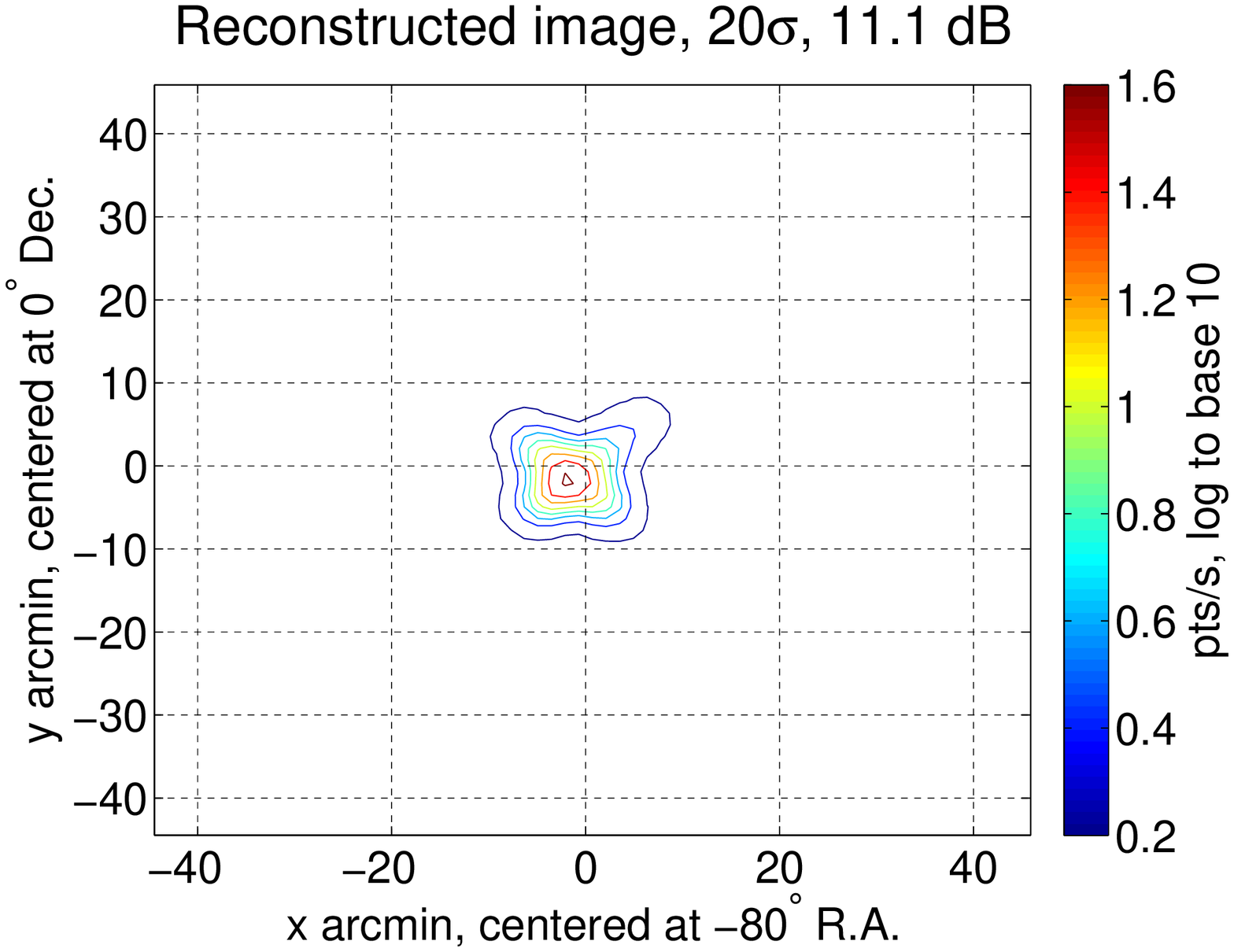}
\includegraphics[width=0.48\linewidth]{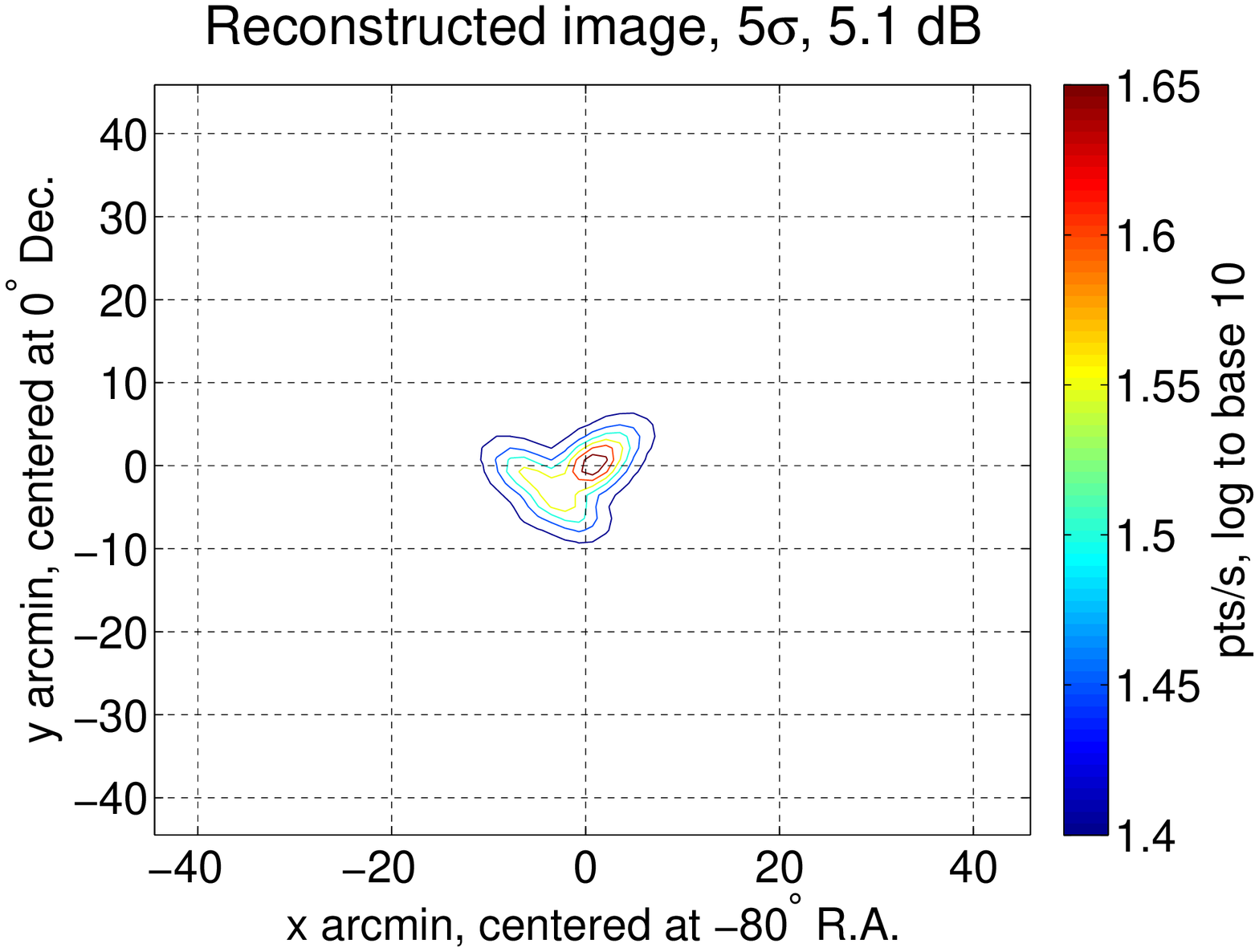}\\
\includegraphics[width=0.48\linewidth]{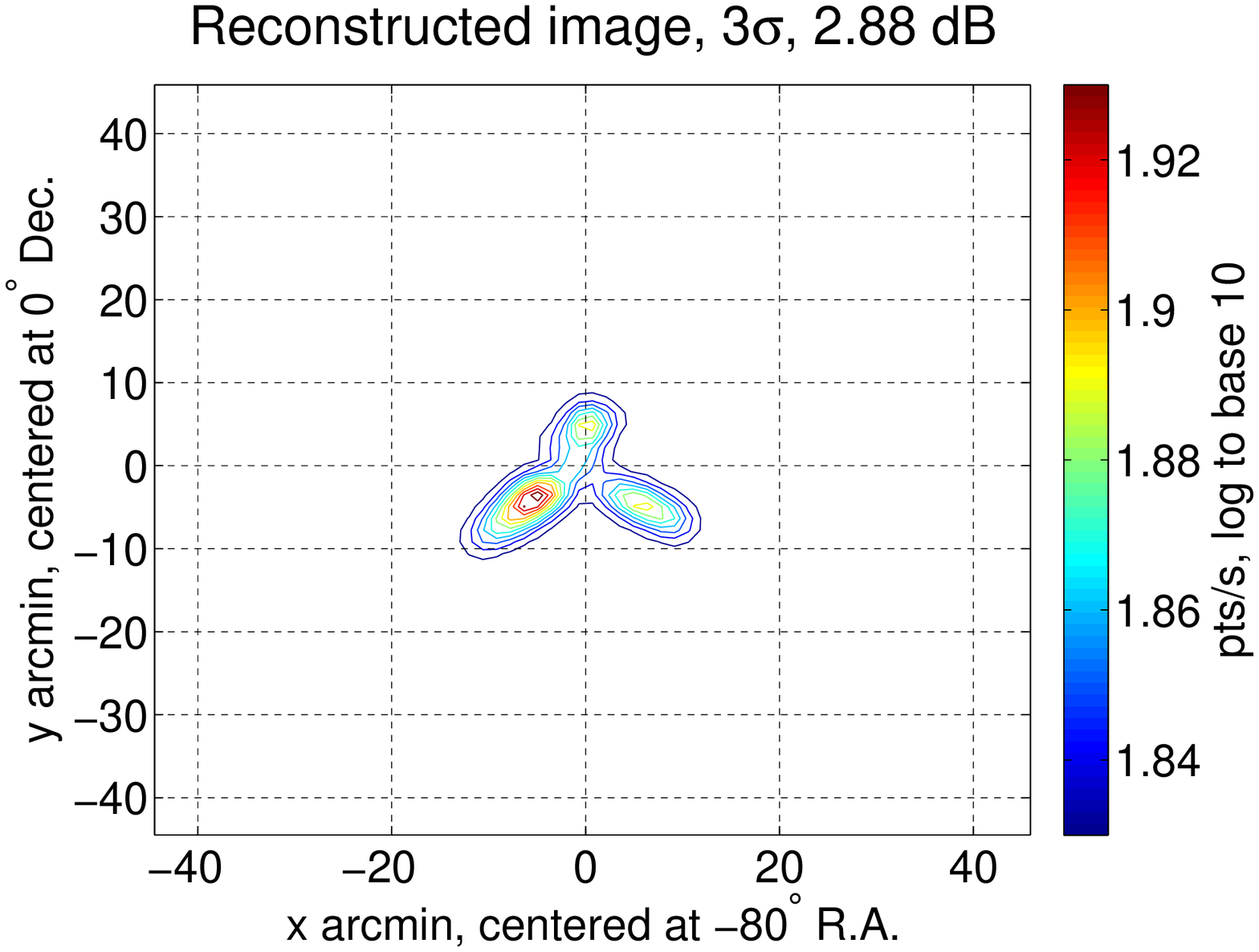}
\includegraphics[width=0.48\linewidth]{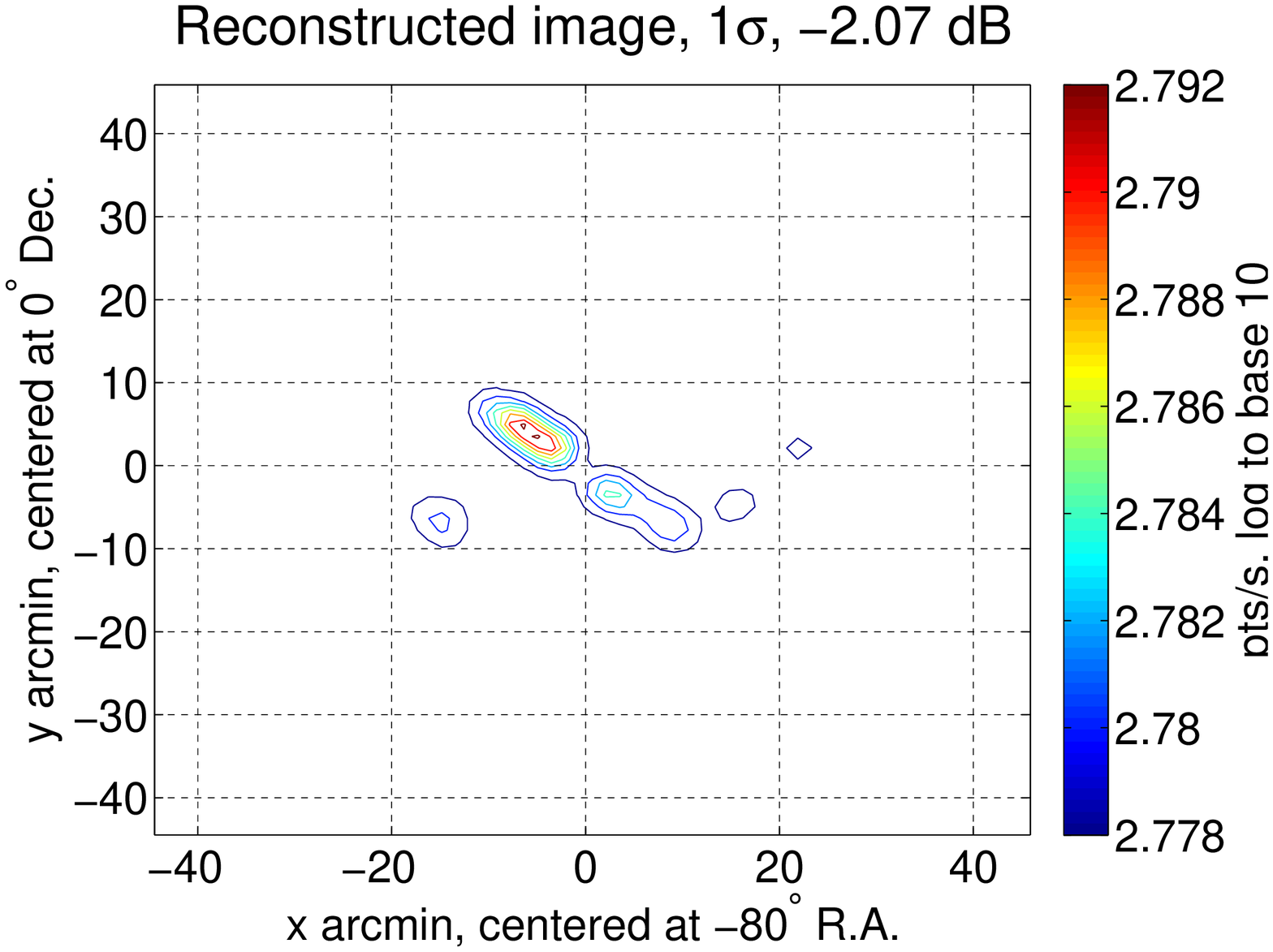}
\caption{\textbf{The upper 4 images:}
Observed data of point sources with $20 \sigma$, $5 \sigma$, $3 \sigma$ and $1 \sigma$ significances.
The SNRs are $11.1\;\mathrm{dB}$, $5.1\;\mathrm{dB}$, $2.88\;\mathrm{dB}$ and $-2.07\;\mathrm{dB}$ respectively.
\textbf{The lower 4 contour diagrams:}
Reconstructed images of the point sources.}
\label{fg-std-pts}
\end{figure}

To test the resolving ability and reliability of DD method for HXMT all-sky survey data we set up a point source as well as a uniform background so that the significance of the source is $k \sigma$, where $\sigma$ is the standard deviation of the fluctuation of the background.
We simulated the observed data and the reconstructions for all the set-ups, as shown in Fig. \ref{fg-std-pts}.
As we can see in Fig. \ref{fg-std-pts} bright point sources with significance more than $5 \sigma$ can be reconstructed accurately.
While reconstruction of faint point sources with significance less than $3 \sigma$ is affected by the noise.
Resolution better than $5\;\mathrm{arcmin}$ and positioning accuracy better than $2\;\mathrm{arcmin}$ can be achieved with our method.

\section{Conclusions}
In this article we present the approximation of the kernel function for image reconstruction from HXMT data and the pixelization scheme optimized for this accelerated DD method.
The pixels are nearly squares with the same size to make up a grid suitable for convolution.
However due to the spherical topology it is impossible to tessellate the spherical surface without gap or overlap with perfect spherical squares of the same size.
The error of the sizes of the pixels are propagated to the results of numerical integrations.
The distortions of the pixels cause the error of distance between any two pixels, which are propagated to the numerical convolutions eventually.
These errors will not be accumulated or magnified through iterations because each iteration of DD method is based on the modulation equation directly\citep{li1993}.
The efficiency of the spherical tessellation is not optimal, since two adjacent tessellae on the same latitudes overlap each other.
This is a trade-off for better regularity of the pixelization scheme (with better uniformity and better orthogonality).
Extra computational costs are brought by the overlaps, as shown in Table \ref{tb-redundancy}.
\begin{table}[htbp]
\centering
\begin{tabular}{r|r}
Tessella size                    & Relative extra computational costs \\
\hline
$22.5^\circ \times 22.5^\circ$   & $20.3\%$ \\
$15^\circ \times 15^\circ$       & $14.5\%$ \\
$11.25^\circ \times 11.25^\circ$ & $11.1\%$ \\
$9^\circ \times 9^\circ$         & $10.0\%$ \\
$5.625^\circ \times 5.625^\circ$ & $ 6.0\%$
\end{tabular}
\caption{Extra computational costs due to the redundancy of the spherical pixelization}
\label{tb-redundancy}
\end{table}

With simulated data we demonstrate the method we proposed works for both equatorial data and data observed in high latitude area.
We can obtain a reconstructed image from the given data in several minutes on ordinary desktop PCs.

Since we are focused on the acceleration of the DD method, some other topics are not discussed, such as noise suppression, treatment of faint and extended sources, which are also necessary for image reconstruction for HXMT data.
For example, in Fig. \ref{fg-obs-ddm} we adjust the SNR of the observed images so that images that with the model images can be achieved although a number of the sources in the INTEGRAL 7-year catalog are actually too faint for a single phase of the HXMT all-sky survey.
In addition we use known backgrounds as background constraints in the DD method, which is usually too idealized to realize in practice.

The acceleration technique we proposed in this article is necessary and sufficient for our future work. It serves as a tunable underlying engine of the flexible DD method.

\appendix
\section{Quaternion and related operations}
\label{app-quat}
In 3-D Cartesian coordinate system a \textbf{quaternion} is defined as
\begin{equation}
\boldsymbol{q} = a + b\mathrm{i} + c\mathrm{j} + d\mathrm{k}\text{,}
\end{equation}
where $a$ is its scalar part and $b\mathrm{i} + c\mathrm{j} + d\mathrm{k}$ is its vector part.

A quaternion $\boldsymbol{q} = a + b\mathrm{i} + c\mathrm{j} + d\mathrm{k}$ is a \textbf{unit quaternion} if and only if $a^2+b^2+c^2+c^2=1$.
A unit quaternion is used to formulate a rotation performed on a rigid body in 3-D space.
Given a unit quaternion
\begin{equation}
\boldsymbol{q} = a + b\mathrm{i} + c\mathrm{j} + d\mathrm{k} = \cos\frac{\alpha}{2} + \sin\frac{\alpha}{2}(\cos\theta\cos\phi\mathrm{i} + \cos\theta\sin\phi\mathrm{j} + \sin\theta\mathrm{k})\text{,}
\end{equation}
$\boldsymbol{q}$ then indicates the right-handed rotation around vector $\Bigr(\begin{smallmatrix}\cos\theta\cos\phi\\
\cos\theta\sin\phi\\
\sin\theta\end{smallmatrix}\Bigl) = \cos\theta\cos\phi\mathrm{i} + 
\cos\theta\sin\phi\mathrm{j} + 
\sin\theta\mathrm{k}$ by angle $\alpha$.

The \textbf{conjugate of quaternion} $\boldsymbol{q} = a + b\mathrm{i} + c\mathrm{j} + d\mathrm{k}$ is defined as
\begin{equation}
\bar{\boldsymbol{q}}=a - b\mathrm{i} - c\mathrm{j} - d\mathrm{k}\text{.}
\end{equation}
If the quaternion $\boldsymbol{q}$ is unit then its conjugate $\bar{\boldsymbol{q}}$ is equivalent to its \textbf{inverse} $\boldsymbol{q}^{-1}$.

Given the equation of basis elements $\mathrm{i}$, $\mathrm{j}$ and $\mathrm{k}$ of quaternions as well as 3-D vectors
\begin{equation}
\mathrm{i}^2=\mathrm{j}^2=\mathrm{k}^2=\mathrm{ijk}=-1\text{,}
\end{equation}
the multiplication of quaternions and 3-D vectors can be calculated distributively.

Let $\boldsymbol{q}$ and $\boldsymbol{v}$ be a unit quaternion and an arbitrary 3-D vector respectively, hence the multiplication $\boldsymbol{q}\boldsymbol{v}\boldsymbol{q}^{-1}$ yields the vector we obtain on rotating $\boldsymbol{v}$ as $\boldsymbol{q}$ indicates.

\section{Point-in-square-on-sphere problem and Spherical-ray-casting algorithm}
\label{app-pisos}
The point-in-square-on-sphere (PISOS) problem asks whether a given point on the spherical surface lies inside, outside or on the boundary of a spherical square.
A spherical square is a regular spherical quadrilateral which has four equal sides and four equal angles.
Each side is the arc of great circle passes two adjacent vertices while each angle is formed by the tangents of two adjacent sides.

The PISOS problem arises when we try to decide which observed data or pixel lies in which sky region.
This problem is a special case of PIP (Point-in-polygon) problem in computational geometry, which can be tackled by the ray-casting algorithm.
According to the algorithm, one can find whether a given point is inside or outside a polygon by testing how many times a ray, starting from any known point inside the polygon and going towards the given point, intersects the edges of the polygon.
If the number of the intersections is odd the point is outside while it is inside if even.

We designed a more specific method, the spherical-ray-casting algorithm, to solve the PISOS problem. Assuming the spherical square is smaller than a semispherical surface, the PISOS problem reduces into the following problems:
\begin{enumerate}
\item Find intersections of two given great circles on a unit sphere.
Either great circle is defined by two points on the unit spherical surface.

This problem is reduced into finding the normal vector of a plane defined by two points on a unit spherical surface as well as the orgin.
For example, let $\boldsymbol{p}_1$, $\boldsymbol{p}_2$, $\boldsymbol{p}_3$ and $\boldsymbol{p}_4$ be points on the unit sphere $x^2+y^2+z^2=1$, where $\boldsymbol{n}_1$ is the normal vector of the plane $\boldsymbol{p}_1\boldsymbol{O}\boldsymbol{p}_2$ and $\boldsymbol{n}_2$ of $\boldsymbol{p}_3\boldsymbol{O}\boldsymbol{p}_4$, i.e., $\boldsymbol{n}_1 = \frac{\boldsymbol{p}_1 \times \boldsymbol{p}_2}{|\boldsymbol{p}_1 \times \boldsymbol{p}_2|}$ and $\boldsymbol{n}_2 = \frac{\boldsymbol{p}_3 \times \boldsymbol{p}_4}{|\boldsymbol{p}_3 \times \boldsymbol{p}_4|}$.
If $\boldsymbol{q}$ is the intersection of plane $\boldsymbol{p}_1\boldsymbol{O}\boldsymbol{p}_2$ and $\boldsymbol{p}_3\boldsymbol{O}\boldsymbol{p}_4$, vector $\boldsymbol{q}$ is perpendicular to both $\boldsymbol{n}_1$ and $\boldsymbol{n}_2$ thus any vector on the plane $\boldsymbol{n}_1\boldsymbol{O}\boldsymbol{n}_2$, therefore $\boldsymbol{q}$ is the normal vector of this plane, i.e., $\boldsymbol{q} = \frac{\boldsymbol{n}_1 \times \boldsymbol{n}_2}{|\boldsymbol{n}_1 \times \boldsymbol{n}_2|}$.

Let $\theta_1$, $\phi_1$ and $\theta_2$, $\phi_2$ be the altitudes and azimuthal angles of two points on the unit sphere while $\theta_n$ and $\phi_n$ be those of their normal vector.
We have
\begin{equation}
\left\{\begin{aligned}
&\cos\theta_n\cos\phi_n\cos\theta_1\cos\phi_1 + \cos\theta_n\sin\phi_n\cos\theta_1\sin\phi_1 + \sin\theta_n\sin\theta_1 = 0\\
&\cos\theta_n\cos\phi_n\cos\theta_2\cos\phi_2 + \cos\theta_n\sin\phi_n\cos\theta_2\sin\phi_2 + \sin\theta_n\sin\theta_2 = 0
\end{aligned}\right.\text{,}
\end{equation}
hence,
\begin{equation}
\left\{\begin{aligned}
&D=\sqrt{\cos^2\theta_1\sin^2\theta_2+\sin^2\theta_1\cos^2\theta_2-2\cos\theta_1\cos\theta_2\sin\theta_1\sin\theta_2\cos(\phi_1-\phi_2)}\\
&\tan\theta_n = \frac{\cos\theta_1\cos\theta_2\sin(\phi_2-\phi_1)}{D}\\
&\sin\phi_n = \frac{\sin\theta_1\cos\theta_2\cos\phi_2 - \cos\theta_1\sin\theta_2\cos\phi_1}{D}\\
&\cos\phi_n = \frac{\cos\theta_1\sin\theta_2\sin\phi_1 - \sin\theta_1\cos\theta_2\sin\phi_2}{D}
\end{aligned}\right.\text{,}
\end{equation}
where $D$ is an auxiliary quantity.

\item Find whether a given minor arc of great circle contains a point on the same great circle.
Let $\boldsymbol{p}_0$ and $\boldsymbol{p}_1$ be the endpoints of an minor arc of great circle and let $\boldsymbol{q}$ be a point on the great circle.
The point $\boldsymbol{q}$ is on the given arc if and only if the sum of geodesic distances from $\boldsymbol{q}$ to $\boldsymbol{p}_0$ and to $\boldsymbol{p}_1$ equals to the length of the arc.
\end{enumerate}

\section*{Acknowledgements}
This work was supported by the National Natural Science Foundation of China (NSFC) under grants No. 11173038 and No. 11103022, and also by Tsinghua University Initiative Scientific Research Program under grant No. 20111081102.

\bibliographystyle{raa}
\bibliography{main}
\end{document}